\documentclass[12pt]{article}
\usepackage{amssymb,latexsym,amsfonts}
\newcommand{\bm}[1]%
 {\mbox{\boldmath $#1$}}
\newcommand{\half}{{\scriptstyle{{1\over 2}}}}

\def\mmin{{\scriptstyle{-}}}
\def\beq{\begin{equation}}
\def\eeq{\end{equation}}
\def\bea{\begin{array}}
\def\eea{\end{array}}
\def\beqa{\begin{eqnarray}}
\def\eeqa{\end{eqnarray}}
\newcommand{\refeq}[1]{\mbox{eq.~(\ref{eq:#1})}}
\def\myIm{{\Im }}

\def\myRe{{\Re }}

\def\u1{{U(1)}}
\def\su2{{SU(2)}}
\def\sun{{SU(n)}}
\def\etp{{\frac{1}{2\pi}}}

\def\ie{{i.e.\/}}\relax
\def\eg{{e.g.\/}}\relax
\relax
\def\diag{{\rm diag}}
\def\w{{\rm w}}
\def\e{{\rm e}}
\def\omw{{\w}}
\def\Omw{{W}}
\def\mon{{\rm m}}
\def\ca{{\rm c}}
\def\rel{{\rm rel}}
\def\ttau{{\tau}}
\newcommand{\complex}{\mathbb{C}}
\newcommand{\qu}{\mathbb{H}}
\newcommand{\re}{\mathbb{R}}
\newcommand{\zahlen}{\mathbb{Z}}

\def\cT{{\cal{T}}}
\def\cO{{\cal{O}}}
\def\cG{{\cal{G}}}

\def\cP{{\cal{P}}}
\def\cM{{\cal{M}}}
\def\cN{{\cal{N}}}

\def\cW{{\vec{\cal W}}}
\def\cY{\hat{\cal{Y}}}
\def\cC{\hat{\cal{C}}}
\def\ddz{{\frac{d}{dz}}}
\def\Tr{{\rm Tr}} 
\def\tr{{\rm tr}}

\def\pl{{{\cal P}_\infty}}
\def\mma{{\cal M}}
\def\hkom{{\vec \omega}}
\def\sd{self\-du\-al}
\def\asd{an\-ti\--\-self\-du\-al}
\def\hk{hyper\-K\"ahler}
\def\hkqc{\hk\ quotient con\-struc\-tion}
\def\thk{toric \hk}
\def\ms{{moduli\ space}}
\def\g{{\mathfrak g}}
\def\c{{\mathfrak c}}

\def\i{{\bf i}}
\def\j{{\bf j}}
\def\k{{\bf k}}
\def\mmu{{\vec \mu}}
\def\r{{\vec r}}
\def\K{K\"ahler}

\def\pl{{{\cal P}_\infty}}
\def\plo{{{\cal P}_\infty^0}}

\def\oo{{(1,1,\ldots,1)}}
\begin{document}
\hfill INLO-PUB-12/98
\vskip1cm
\begin{center}
{\LARGE{\bf{\underline{Instantons, Monopoles and}}}}\\
{\LARGE{\bf{\underline{Toric HyperK\"ahler Manifolds}}}}\\
\vspace{7mm}
{\large Thomas C. Kraan} \\
\vspace{5mm}
Instituut-Lorentz for Theoretical Physics, University of Leiden,\\
PO Box 9506, NL-2300 RA Leiden, The Netherlands.
\end{center}
\vspace*{5mm}{\narrower\narrower{\noindent
\underline{Abstract:} 

In this paper, the metric on the moduli space of the $k=1$ $SU(n)$ periodic instanton -or caloron- with arbitrary gauge holonomy at spatial infinity is explicitly constructed. 
The metric is toric \hk\ and of the form conjectured by Lee and Yi. 
The torus coordinates describe the residual $U(1)^{n-1}$ gauge invariance and the 
temporal position of the caloron 
and can also be viewed as the phases of $n$ monopoles that constitute the caloron. 
The $(1,1,\ldots,1)$ monopole is obtained as a limit of the caloron. 
The calculation is performed on the space of Nahm data, 
which is 
justified by proving the isometric 
property of the Nahm construction for the cases considered. 
An alternative construction using the \hk\ quotient is also presented. 
The effect of massless monopoles is briefly discussed.
}\par}

\section{Introduction}
Moduli spaces of instantons \cite{ADHM} and Bogomol'nyi-Prasad-Sommerfield (BPS) monopoles \cite{BPS} have been subject to long-time investigation. 
The moduli space, quotient of the set of self-dual gauge connections by the group of gauge transformations,
is a subset of the configuration space and its geometry therefore reflects physical properties of the system. 

In this paper periodic instantons \cite{HarShe} on $\re^3 \times S^1$, or calorons, are studied for gauge group $SU(n)$. 
Calorons are composed out of elementary BPS monopoles \cite{LeeYi1}, 
as is seen from the action density \cite{MC}. 
This becomes clear for small compactification lengths when the constituents are far apart. 
In particular, removing one of the monopoles to spatial infinity turns the $k=1$ caloron into a BPS $SU(n)$ monopole. 
In contrast, the situation of all monopoles nearly coalescing -in appropriate units corresponding to an infinite compactification length- gives back the ordinary instanton on $\re^4$. 
These various aspects are respected by the corresponding limits in the metric.
The form of the metric was conjectured by Lee and Yi \cite{LeeYi1}, using considerations of D-brane constructions and asymptotic monopole interactions. 
This paper addresses the explicit calculation of the metric for the caloron moduli space and its limits. 

Metric properties of moduli spaces of \sd\ connections play an important role in the study of non-perturbative effects of gauge theories.
For instantons the metric appears through the bosonic zero modes in the background of the charge one $SU(2)$ instanton in a calculation to study its physical effects \cite{Hoo}.
The scattering of monopoles can be described as the geodesic motion on the moduli space \cite{ManRem}, relating the metric to the Lagrangian of the interacting monopole system \cite{ManLag}.

The metrics on these moduli spaces are \hk\ \cite{HKLR}. 
This property derives formally from the nature of 
the \sd ity equations themselves \cite{AtiHit, DonKro}. 
It also appears in the Atiyah-Drinfeld-Hitchin-Manin (ADHM) construction of instantons of higher charge, as well as in the Nahm construction for monopoles as a \hk\ structure on the space of data \cite{DonIns, DonMon}.
The Nahm formalism first appeared as a generalisation 
of the ADHM construction to construct the BPS monopole \cite{Nahm}.
In its extension to selfdual monopoles for arbitrary group and charge \cite{NahAll, HurMur}, the Nahm data in terms of which the monopole is obtained can be constructed in terms of the Weyl zero modes 
in the background of the monopole. 
A similar scheme was set up for the caloron \cite{NahMonCal, GarMur}, which up to very 
recently \cite{PLB,NPB,LeeLu} had not resulted in explicit solutions. 
This reciprocity idea could be applied to instantons on $\re^4$ as well \cite{CorGod}.
Extended to the four-torus $T^4$, the involutive property of the Nahm transformation preserves the metric and \hk\ structure \cite{BraBaa}. 
These ideas fit in a programme of studying the Nahm transformation 
on generalised tori $M = \re^4/H$, where $H$ is the isometry group of the \sd\ connection. 
The calorons correspond to $M = \re^3 \times S^1, H = \zahlen$.
This compactification provides a smooth interpolation between instantons and monopoles, adding to the understanding of both objects and the formalism to study them.

The incorporation of both instanton and monopole-like aspects by calorons is read off from the topological characteristics of \sd\ gauge connections $A_\mu dx_\mu$ on $\re^3 \times S^1$ \cite{GroPisYaf}.
These are related to the properties of the vacuum which the solution necessarily approaches at spatial infinity in order for the action to be finite.
The homotopy class of the gauge 
transformation connecting the vacuum at infinity 
with the connection near the origin gives the instanton number $k \in \pi_3(SU(n)) = \zahlen$. 
The vacuum itself can be nontrivial, due to the non-trivial topology of the asymptotic boundary of the base manifold $S^2 \times S^1$. 
This leads to extra labels for the solution which are studied in terms of the gauge holonomy ${\cal P} (\vec x)$ along $S^1$. 
In the periodic gauge ($A_\mu(\vec x,x_0+\cT)=A_\mu(\vec x,x_0)$), ${\cal P}(\vec x)$ is defined as
\beq
{\cal P} ( \vec x) = P \exp( \int_0^{\cal T} A_0(\vec x, x_0) d x_0),
\eeq
where $P$ denotes path ordering and $\cT$ the circumference of $S^1$, which we set $1$. 
In a zero curvature background, continuous deformations 
of the loop do not affect ${\cal P}(\vec x)$. 
Its eigenvalues at spatial infinity are topological invariants. 
Therefore, the gauge holonomy at infinity is diagonal up to an $\hat x$ dependent gauge transformation $V$ 
\beq
\lim_{|\vec x|\rightarrow\infty}{\cal P}(\vec x)=\pl=V\plo V^{-1},\quad\plo
=\exp[2\pi i{\rm diag}(\mu_1,\ldots,\mu_n)].
\eeq
The eigenvalues can be ordered such that
\beq
\mu_1<\ldots<\mu_n<\mu_{n\!+\!1}\equiv\mu_1+1,\quad\sum_{m=1}^n\mu_m=0,
\eeq
using the gauge symmetry and assuming maximal symmetry breaking for the moment. For later use, we define $\nu_m=\mu_{m\!+\!1}-\mu_m$, related to the mass of the $m^{\rm th}$ constituent 
monopole. 
Asymptotically, 
\beq
A_0=2\pi i\,{\rm diag}(\mu_1,\cdots,\mu_n)-i\,{\rm diag}(k_1,\cdots,k_n)/(2r)+
\cO(r^{-2}),\quad\sum_i k_i=0,\label{eq:asa0}
\eeq
up to the gauge transformation $V(\hat x)$ that induces a map from $S^2$ to $SU(n)/H_\infty$, with $H_\infty$ the isotropy group of $\exp[2\pi i\,{\rm diag}
(\mu_1,\cdots,\mu_n)]$. 
The maps $V(\hat x)\rightarrow SU(n)
/H_\infty$ are classified according to the fundamental group of $H_\infty$. 
Generically, $H_\infty$ consists of several $U(1)$ and $SU(N),\,N >1$ 
subgroups. 
Each $U(1)$ gives rise to a monopole winding number, related to 
the integers $k_i$. 
The enhanced residual gauge symmetry described by the $SU(N)$ subgroups arises when there is non-maximal symmetry breaking, $\nu_m = \mu_{m+1} - \mu_m=0$ for some value(s) of $m$, giving rise to massless constituent monopoles.
A non-trivial value of $\cal P_\infty$ breaks the gauge symmetry. 
This makes calorons very similar to BPS monopoles, \cite{Nahm, NahAll, HouSut} which fit in the above classification as $S^1$ invariant \sd\ connections, classified according to the magnetic charges $(m_1, \ldots, m_{n-1})$, where $m_i = k_1 + \ldots+ k_i$.
The $k=1$ $SU(n)$ caloron studied in this paper has no magnetic charges, 
and its only nontrivial topological labels are the instanton number $k=1$ and the eigenvalues $\mu_m$ of the holonomy. 

The explicit computation of the metrics in this paper is 
based on the isometric property of the Nahm 
transformation, known to hold for instantons on $\re^4$ and $T^4$,
as well as for certain types of BPS monopoles \cite{Nak}.
It is believed to hold generally. 
For most situations considered in this paper,
an explicit proof seems not to be present in the literature,
and will be given here.
This allows for a determination of the metric on the moduli space of Nahm data. 
For monopoles, such a calculation was first done in \cite{Con} showing
that the 
metric of the $(1,1)$ data is a Taub-NUT space with 
positive mass parameter.
Considerations based on asymptotic monopole interactions \cite{GibManBPS}
reproduced this result \cite{su3}. 
For the $(1,1,\ldots,1)$ 
monopole a similar equivalence was found \cite{LeeWeiYi1, MurNot}.
All these metrics are of so-called \thk\ type \cite{PedPoo, GGPT}, 
and can be efficiently obtained as metrics on \hk\ 
quotients \cite{GibRycGot}. An explicit calculation 
of the $k=1$ $SU(2)$ caloron is extended here to $SU(n)$, 
generalising the techniques 
in \cite{PLB, NPB}.
An alternative derivation using the \hk\ quotient will also be given. 
There we will greatly benefit from the formalism in \cite{MurNot, GibRycGot}, due to the similarity
between the caloron and monopole Nahm data.

The outline of this paper is as follows. 
In section 2, some aspects of \hk\ manifolds are presented, 
mostly to fix notation and to give some identities used throughout.
Crucial in the ability to handle the caloron is that the infinite
matrices of the ADHM construction are converted by Fourier transformation
to functions on $S^1$. 
This translates ADHM to the Nahm formulation
and allows one to keep track of crucial delta-function singularities.
In section 3, to define notation, 
we summarise the ADHMN formalism for 
calorons as developed in refs. \cite{PLB, NPB, LeeYi2} based on 
the ADHM construction for 
instantons, rather than following \cite{NahMonCal, GarMur}.
The caloron metric is calculated in section 4. 
The instanton and monopole limits of the caloron are discussed in section 5. 
A unified description of instantons, calorons and monopoles is thus achieved. 
Other aspects of the caloron,
among which the effect of massless constituents, 
are commented on in the discussion.
The appendix contains some technicalities on the $\oo$ monopole.

\section{Preliminaries} \label{sec:prel}
Manifolds with metric $g$ are \hk\ if they have three independent complex structures $I,J,K$ that satisfy the quaternion algebra, $ IJ = - JI = K$ and cyclic, whose associated \K\ forms $\omega^{I}(\cdot, \cdot) = g( \cdot, I \cdot)$, $\omega^{J}(\cdot, \cdot) = g( \cdot, J \cdot)$, $\omega^{K}(\cdot, \cdot)= g( \cdot, K \cdot)$ are closed.
As will be outlined in section \ref{sec:modsp}, the moduli spaces of \sd\ connections inherit their \hk\ property from the \hk\ structure of the base space manifold $M = \re^4/H$, where $H= \emptyset, \zahlen, \re$ for instantons, calorons
and monopoles respectively.
The position coordinate on $\re^4$ will be denoted as a quaternion, $x = x_\mu \sigma_\mu$. Here the unit quaternions are defined as $\sigma_\mu = (1_2, -i \vec \tau)=( 1, \i,\j,\k)$ and $\bar \sigma_\mu = ( 1_2, i \vec\tau)$, with $ \i \j = - \j \i = \k $ and $\tau$ the Pauli matrices. 
We introduce the \sd, resp. \asd\ quaternionic tensors \cite{Hoo} $\eta_{\mu\nu} \equiv \eta^i_{\mu\nu} \sigma_i \equiv \half(\sigma_{\mu} \bar \sigma_{\nu}-\sigma_{\nu} \bar \sigma_{\mu}) $ and $\bar\eta_{\mu\nu} \equiv \bar\eta^i_{\mu\nu} \sigma_i\equiv\half(\bar\sigma_{\mu} \sigma_{\nu}-\bar\sigma_{\nu} \sigma_{\mu})$, and $\epsilon_{0123} = 1$.
Identifying the tangent space to $\qu = \re^4$ with the vector space itself, 
the complex structures act on $x$ as right multiplication with $- \i, -\j, -\k$, such that $(I,J,K)_{\mu\nu} = \bar\eta^{1,2,3}_{\mu\nu}$. 
It is convenient to combine the metric and \K\ forms into one quaternion,
\beq
(g, \hkom )= g \sigma_0+ \hkom \cdot \vec \sigma .
\eeq
This implies for $\re^4$,
\beqa 
&(g, \hkom )= d \bar x \otimes d x,&\nonumber\\ 
&g = ds^2 = (d x_\mu)^2, &\nonumber\\
& \hkom \cdot \vec\sigma = \half d\bar x \wedge d x = \half \bar\eta_{\mu\nu} d x_\mu \wedge d x_\nu = d x_0 \wedge d \vec x-\half d \vec x \wedge d\vec x &\label{eq:defs}.
\eeqa
Here, $(d \vec a \wedge d \vec b )^i = \epsilon_{ijk} d a^j \wedge d b^k$.
 One extends to
$\qu^N$ by replacing $d \bar x$ in \refeq{defs} by $d x^\dagger = d\bar x^t$. 

Many examples of \hk\ manifolds emerge as \hk\ quotients \cite{HKLR}. 
Consider a \hk\ manifold $\mma$ acted upon freely by a group $G$ (with algebra $\g$) of isometries, $L_X g = 0$, $L$ denoting the Lie derivative and $X \in \g$. 
When $G$ preserves the complex structures,
$
L_X \hkom=0,
$
the isometries are called triholomorphic and the moment map $\mmu: \mma \rightarrow \g^* \otimes \re^3$ can be defined as
$X_\mu \hkom_{\mu\nu} = \partial_\nu \mmu^X$. The manifold $\mmu^{-1}({\vec\c})/G$, with $\vec\c \in \re^3 \otimes Z_\g$ ($Z_\g$ the center of $\g^*$) obtained by taking the quotient of the level set $\mmu^{-1}(\vec\c)$ by $G$ is then \hk\ itself. 
Isometries commuting with $G$ descend to the quotient. When they are also triholomorphic, this property is preserved. 

The relevant example is provided by the moduli space of ADHM data in the construction of charge $k$ instantons on $\re^4$ for gauge group $SU(n)$ \cite{ADHM, CorGod}. 
The caloron will be constructed using an infinite-dimensional version of the ADHM construction which we therefore review here, to establish conventions. 
One considers the set $\hat {\cal A}$ of matrices 
\beq
\Delta = \left( \bea{c} \lambda\\ B \eea \right),
\eeq
with $\lambda \in \complex^{n, 2 k}$ and
the $2k \times 2k$ dimensional matrix $B = B_\mu \otimes \sigma_\mu$, where $B_\mu$ are $k \times k$ dimensional hermitean matrices. 
With metric and \K\ forms on $\hat{\cal A}$ defined as 
\beq
(g, \hkom) = \Tr \left( dB^\dagger \otimes dB + 2 d\lambda^\dagger \otimes d\lambda \right), \label{eq:mkfadhm}
\eeq
$\hat{\cal A}$ is \hk .
The $U(k)$ transformations
\beq
\lambda \rightarrow \lambda T^\dagger, \quad B_\mu \rightarrow T B_\mu T^\dagger
,\quad T \in U(k), \label{eq:symmadhm}
\eeq
leave $(g, \hkom)$ in \refeq{mkfadhm} invariant and therefore form a group of triholomorphic isometries of $\hat{\cal A}$.
The associated moment map reads ($\tr_2$ denoting the trace associated with quaternions)
\beq
\mmu = \half\tr_2 \left[ \left( B^\dagger B+\lambda^\dagger\lambda\right)\vec{\bar\sigma} \right].
\eeq 
Its zero set $\mmu^{-1}(0)$ is formed by the solutions to the ADHM constraint
\beq
\bar\eta_{\mu\nu} B_\mu B_\nu + \half \tau_a \tr_2( \tau_a \lambda^\dagger \lambda) = 0.
\eeq
The instanton gauge connection corresponding to a solution to $\Delta \in \mmu^{-1}(0)$ is obtained as
\beq
A_\mu(x)=v^\dagger(x)\partial_\mu v(x), \label{eq:aadhm}
\eeq
in terms of the $(2 k + n) \times n$ dimensional complex matrix $v(x)$ containing the normalised zero modes of $\Delta^\dagger(x) = \Delta^\dagger - x^\dagger b^\dagger$, where $b^\dagger = (0, 1_k)$.
For $A_\mu$ to be an $SU(n)$ gauge potential, $B^\dagger B + \lambda^\dagger \lambda$ should be invertible, implying the existence of a $k\times k$ dimensional hermitean matrix $f_x$ commuting with the quaternions,
\beq
\Delta^\dagger(x) \Delta(x) = \sigma_0 \otimes f^{-1}_x. \label{eq:df}
\eeq
This matrix features in the expression for the curvature, 
\beq
F_{\mu\nu} = 2 v^\dagger(x) b\eta_{\mu\nu} f_x b^\dagger v(x),
\eeq
showing it to be self-dual. 
It also appears in the formula for the
 action density \cite{Osb},
\beq
\Tr F_{\mu\nu}^2(x)=-\partial_\mu^2\partial_\nu^2\log\det f_x \label{eq:ad}
\eeq
from which it follows that the topological charge is $k$, because of
the asymptotic behaviour 
\beq
f_x = 1_k/x^2, \quad x^2 \rightarrow \infty \label{eq:fadhmas}.
\eeq
Thus it is shown that an element 
$\Delta \in \mmu^{-1}(0)$ corresponds to a charge 
$k$ instanton solution. 
The gauge connection (\ref{eq:aadhm}) is
not affected by the $ U(k)$ transformations (\ref{eq:symmadhm}), which therefore have to be divided out to obtain the instanton
moduli space $\mmu^{-1}(0)/ U(k)$ (its isometry with the moduli space of instantons is discussed later). 
This reduces the dimension of the instanton moduli space to $4 k n$. As it is a \hk\ quotient, this space is 
\hk\ \cite{DonIns, DonKro}. 
Global gauge transformations of 
the instanton which are included as moduli, 
are realised by the action
\beq
\lambda \rightarrow g \lambda, 
\quad g \in SU(n), \label{eq:global}
\eeq
which is a triholomorphic isometry, as follows from \refeq{mkfadhm}. 
As $SU(n)$ acts on the left, it commutes with $U(k)$ acting on the right. Therefore, $SU(n)$ descends as a group of triholomorphic isometries to the moduli space of ADHM data, the \hk\ quotient
$\mmu^{-1}(0)/U(k)$, reflecting the gauge symmetry of the
instanton solution.

At this place we recall a frequently 
used $U(1)$ fibration over $\re^3$, physically interpreted as a monopole phase and position. 
It is presented in terms of complex row 2-vectors that feature in the ADHM matrix $\lambda$. 
Specifically, 
for a 2-dimensional complex row vector 
$\varsigma = (\varsigma_1, \varsigma_2)$, 
describing $\re^4$, the metric and K\"ahler forms read 
\beq
(g, \hkom) = d \varsigma^\dagger \otimes d \varsigma, \quad g = \half \tr_2 d\varsigma^\dagger d \varsigma, 
\quad \hkom \cdot \vec\sigma= \half d\varsigma^\dagger \wedge d \varsigma. \label{eq:altmkfo}
\eeq
The complex structures act on 
$\varsigma$ by right multiplication with 
$ -\sigma_i$. 
There is a triholomorphic $U(1)$ isometry with associated moment map
\beq
\varsigma \rightarrow e^{it} \varsigma, \quad \mmu = \half \tr_2 (-i \varsigma^\dagger \varsigma \vec{\bar \sigma}) = \half \vec r.\label{eq:level}
\eeq
The level sets are
$U(1)$ fibres due to the phase ambiguity
in defining $\varsigma$ from $\vec r$,
which becomes more manifest upon introducing new coordinates,
\beq
\varsigma = \varsigma^{0}e^{i \frac{\psi}{2}},\quad
 \psi \in \re/( 4 \pi \zahlen)
\eeq
with for example $\varsigma^{0}_2(\vec r)$ chosen real. A useful identity is
\beq
\half \tr_2 ( \delta \varsigma^\dagger_0 \varsigma_0 - \varsigma_0^\dagger \delta \varsigma_0) = -i |\vec r| \vec\omw(\vec r) \cdot d \vec r \label{eq:ident},
\eeq
where $\vec\omw(\vec r)$ is the vector potential of the abelian Dirac monopole,
\beq
\vec\nabla_{\vec r} \times \vec \omw(\vec r) = \vec\nabla_{\vec r} \frac{1}{|\vec r|}.\label{eq:dirac}
\eeq
In the present form, the Dirac string lies along the positive $z$ axis, other gauges are obtained by allowing for $\r$ dependent phase ambiguities.
In terms of $(\vec r, \psi)$, the metric and K\"ahler forms on $\re^4$ read
\beqa
ds^2 = \frac{1}{4}\left(\frac{1}{|\r|} d \r^2 + |\r| ( d \psi +\vec \omw(\r) \cdot d \r)^2\right), \quad
\vec\omega = \frac{1}{4 } (d \psi + \vec \w(\vec r) \cdot d \vec r ) \wedge d \vec r - \frac{1}{4 r} d \vec r \wedge d\vec r. \label{eq:psirmet}
\eeqa
The $U(1)$ isometry is equivalent to a linear action
\beq
\psi \rightarrow \psi + 2 t, \quad t \in \re/(2 \pi \zahlen).
\eeq

The moduli spaces we will encounter are all so-called toric \hk\ manifolds \cite{PedPoo}. These manifolds have coordinates consisting of $N$ three vectors $\vec x_a \in \re^3, a = 1, \ldots, N$, and $N$ torus variables $\phi_a$, generalising the $U(1)$ in the previous example. Metric and \K\ forms read
\[
g = d \vec x_a \Phi_{a b} \cdot d \vec x_{b} + 
 \left(\frac{d \phi_a}{4 \pi} + \vec\Omega_{ac} \cdot d \vec x_{c}\right) (\Phi^{-1})_{ab} \left(\frac{d \phi_b}{4 \pi} + \vec\Omega_{bd} \cdot d \vec x_{d} \right),
\]
\beq
\hkom = ( \frac{d \phi_a}{4 \pi} + \vec\Omega_{ab} \cdot d \vec x_{b}) \wedge d \vec x_{a} - \half \Phi_{a b} d \vec x_{b} \wedge d \vec x_{a}. \label{eq:thkm}
\eeq
The potentials $\Phi$ and $\vec\Omega$ are $\phi_a$ independent, giving rise to $N$ commuting triholomorphic isometries $\partial/\partial \phi_a$, corresponding to shifts on the torus. 
Closure of the \K\ forms is equivalent to 
\beq
\frac{\partial}{\partial x^i_a} \Omega^j_{bc} - \frac{\partial}{\partial x^j_c} \Omega^i_{ba}
=\epsilon_{ijk} \frac{\partial}{\partial x^k_c} \Phi_{ab},\quad \forall a,b,c,i,j. \label{eq:hkcon}
\eeq
These equations are therefore called \hk\ conditions \cite{PedPoo,GGPT}, and generalise \refeq{dirac}.
The metric in \refeq{thkm} has an $SO(3)$ isometry, acting on the vectors $\vec x_a$, that rotates the complex structures.
Toric \hk\ manifolds are torus bundles over $(\re^3)^N$ \cite{GibManBPS}.
Physically, the $\re^3$ vectors $\vec x_a$ are (relative) constituent monopole positions, whereas the torus describes the phases of the monopoles. In the Lagrangian interpretation of the metric, $\Phi$ and $\vec \Omega$ denote retarded interaction potentials for the constituents \cite{ManLag, GibManBPS} and it was considerations of this kind that led to the conjectures for the metric in \cite{LeeYi1,LeeWeiYi1}. 
 
\section{The ADHM-Nahm formalism} \label{sec:formalism}
We will construct the caloron in the so-called algebraic gauge, related to the periodic gauge by the non-periodic gauge transformation
$g(\vec x, x_0) =V\exp[2 \pi i x_0{\rm diag}(\mu_1,\ldots, \mu_n)]V^{-1}$. In this gauge, the background field $2\pi i\,{\rm diag}(\mu_1,\cdots,\mu_n)$ in \refeq{asa0} is removed and we have the alternative boundary condition, 
\beq
A_\mu(\vec x,x_0+\cT)=\pl A_\mu(\vec x,x_0)\cP^{-1}_\infty.\label{eq:algau}
\eeq
Since in the absence of magnetic windings, $\pl$ can always be gauged to a constant diagonal form, we assume henceforth $\pl = \plo$ without loss of generality. The periodic instanton of charge one is obtained in the algebraic gauge (\ref{eq:algau}) by taking an infinite array of elementary instantons, relatively gauge-rotated by $\cal P_\infty$.

To implement this in the ADHM formalism we take a specific solution for the zero mode vector $v(x)$ in the ADHM construction,
\beq
v(x) = \left(\bea{c} -1_n\\u(x) \eea\right) \varphi^{-\half}(x), \quad u(x) = ( B^\dagger - x^\dagger 1_k)^{-1} \lambda^\dagger, \quad\varphi(x) = 1_n + u^\dagger(x) u(x),
\label{eq:adhmzm}
\eeq
where $\varphi$ is an $n\times n$ positive hermitean matrix.
In terms of these, one obtains
\beq
A_\mu(x) = \varphi^{- \half}(x) ( u^\dagger(x) \partial_\mu u(x)) \varphi^{- \half} (x) + \varphi^\half(x) \partial_\mu \varphi^{-\half}(x).\label{eq:adhmveld}
\eeq
For \refeq{algau} to hold, it is then required that 
\beq
u_{p+1}(x+1) = u_p(x) {\cal P}_\infty^{-1}, \quad p \in \zahlen.
\eeq
This imposes periodicity constraints on the data
\beq
\lambda_{p+1} = \pl \lambda_p, \quad B(x+1)_{p,{p'}} = B(x)_{p-1, {p'}-1},
\eeq
which imply
\beq
\lambda_p = {\cal P}_\infty^p \zeta, \quad B_{p,{p'}} =\sigma_0 \delta_{p,{p'}} + \hat A_{p-{p'},} \quad p,{p'} \in \zahlen.
\eeq
The off-diagonal part $\hat A$ is still to be determined. 
Fourier transformation translates the ADHM formalism to the Nahm language. 
$B$ is cast into a Weyl operator, 
\beqa
&&\sum_{p,{p'} \in \zahlen} B_{p,{p'}}(x) e^{2\pi i(pz-{p'}z')}=\frac{\delta(z-z')}{2\pi i}\hat 
D_x(z'),\quad\hat D_x(z)=\sigma_\mu\hat D_x^\mu(z)=\ddz+\hat A(z)-2\pi ix, 
\nonumber\\ 
&&\hat A(z)=\sigma_\mu\hat A^\mu(z),\quad\hat A^\mu(z)=2\pi i\sum_{p\in \zahlen}
e^{2\pi ipz}\hat A^\mu_{p},
\eeqa
and $\lambda^\dagger \lambda$ into a singularity structure describing the matching conditions for $\hat A(z)$,
\beqa
&&\sum_{p\in \zahlen} e^{-2\pi piz}\lambda_p=\sum_{p\in \zahlen} e^{2\pi ip(\mu_m-z)}P_m\zeta=\hat\lambda
(z),\quad\hat\lambda(z)=\sum_{m \in \zahlen/n \zahlen}\delta(z-\mu_m)P_m\zeta,\\
&&\sum_{p,{p'}\in \zahlen}\lambda^\dagger_p e^{2\pi i(pz-{p'}z')}\lambda^{\phantom{\dagger}}_{p'}=
\delta(z-z')\hat\Lambda(z),\quad\hat\Lambda(z)=\sum_{m \in \zahlen/n \zahlen}\delta(z-\mu_m)
\zeta^\dagger P_m\zeta=\zeta^\dagger\hat\lambda(z).\nonumber
\eeqa
Here we introduced the projection operators $P_m \!=\! e_m e_m^t$, where $e_m$ is the $m^{\rm th}$ unit vector, in terms of which $\pl\! = \!\sum_{m \in \zahlen/n \zahlen} \exp( 2 \pi i \mu_m) P_m$ and $\lambda_p = \sum_{m \in \zahlen/n \zahlen} \exp(2 \pi i p\mu_m) P_m \zeta$. 
The group index $m \in \zahlen/n \zahlen$ is a cyclic variable. We also used that for any two objects $a, b$ of type $a_p = {\cal P}_\infty^p \alpha$, 
 $p \in \zahlen$, 
the Fourier transforms defined as
$\hat a(z) = \sum_{p \in \zahlen} \exp( -2 \pi i p z) a_p $, have the property
\beq
\hat a^{\dagger}(z) \hat b (z') = \delta( z - z') \,\hat a(z)^\dagger \! < \!\hat b\!> = \delta( z - z')  < \!\hat a^\dagger\!> \hat b (z) = \delta(z - z') \sum_{m \in \zahlen/n \zahlen} \delta(z - \mu_m) \alpha^\dagger P_m \beta, \eeq
where $ <\!H\!> \equiv \int_{S^1} H(z) dz$.
The quadratic ADHM constraint translates into
\beq
\half [ \hat D_\mu(z), \hat D_\nu(z)] \bar \eta_{\mu\nu} = 4 \pi^2 \myIm \hat \Lambda(z), \label{eq:aq}
\eeq
where $\myIm$ is introduced to act on a $2\times 2$ matrix as $\myIm W \equiv \half [W - \tau_2 W^t \tau_2]$ ($\myRe W \equiv \half \tr_2 W$). 
We use the $U(1)$ fibration over $\re^3$ (\refeq{level}) to write
\beq
\zeta^\dagger P_m \zeta = \zeta^\dagger_{(m)} \zeta^{\vphantom \dagger}_{(m)} = \frac{1}{2 \pi}(\rho_m +\vec\rho_m\cdot \vec \tau), \quad\rho_m=|\vec\rho_m|\label{eq:rho}.
\eeq 
This leads to the caloron Nahm equation
\beq
\ddz \hat A_j(z) = 2 \pi i \sum_{m \in \zahlen/n \zahlen} \delta(z - \mu_m) \rho^j_m, \label{eq:nahmeq}
\eeq
which is abelian in the $k=1$ situation at hand, see \cite{MC, NahMonCal}. The phase ambiguity in defining $\zeta_{(m)}$ from $\vec\rho_m$ is resolved later.
As integration of \refeq{nahmeq} over $S^1$ gives a constraint on $\zeta$, 
\beq
\sum_{m \in \zahlen/n \zahlen} \vec\rho_m = \pi \tr_2 (\vec \tau \zeta^\dagger \zeta) = \vec 0,
\label{eq:conrho}
\eeq 
we can introduce vectors $\vec y_m, m \in \zahlen/n \zahlen$, such that $\vec\rho_m =\vec y_m - \vec y_{m-1}$. 
The vectors $\vec y_m$ are to be interpreted as the constituent monopole positions. 
We now find for the spacelike components of $\hat A(z)$,
\beq
\hat A_j(z) = 2 \pi i \sum_{m \in \zahlen/n \zahlen} \chi_{[\mu_m,\mu_{m+1}]}(z) y^j_m, 
\eeq
where $\chi_{[\mu_m, \mu_{m+1}]}(z) = 1$ for $z \in [\mu_m, \mu_{m+1}] $ and $0$ elsewhere, extended periodically. 
Note that the Nahm equations determine $\vec y_m$ up to the global $\re^3 \times S^1$ position variable
\beq
\xi = \frac{1}{2 \pi i}\int_{S^1} \hat A (z) dz, \label{eq:com}
\quad 
\vec \xi = \sum_{m \in \zahlen/n \zahlen} \nu_m \vec y_m. 
\eeq
The $T$ symmetry \refeq{symmadhm} in the ADHM construction is mapped to a $U(1)$ gauge symmetry, with gauge group $\hat \cG = \{g(z)| g:z \rightarrow e^{- ih(z)} \in U(1)\}$,
acting as
\beq
\hat A(z) \rightarrow \hat A(z) +i \ddz h(z), \quad\zeta_m \rightarrow \zeta_m e^{ih(\mu_m)}. \label{eq:ga}
\eeq
For calorons, $g(z)$ is periodic and can be used to set $\hat A_0(z)$ to a constant. 
A piecewise linear $U(1)$ gauge function $h(z)$ shifts
the $U(1)$ phase ambiguities in $\zeta_{(m)}$ to $\hat A_0(z)$, 
which thus becomes piecewise constant. 
Therefore, all $4n$ moduli are included in the following solution to
the Nahm equations
\beq
\hat A(z) = 2 \pi i \sum_{m \in \zahlen/n \zahlen} \chi_{[\mu_m, \mu_{m+1}]}(z) ( \frac{\ttau_m}{4 \pi \nu_m} \sigma_0 + \vec y_m \cdot \vec \sigma),
\eeq
where $\ttau = ( \ttau_1, \ldots,\ttau_n)^t$ 
takes values in $\re^n$. 
Using the gauge function
\beq
g(z) = \sum_{m \in \zahlen/n \zahlen} \chi_{[\mu_m, \mu_{m+1}]}(z) \exp(2 \pi i ( z - \mu_m)\frac{k_m}{\nu_m} ),\quad k_m \in \zahlen, \label{eq:intijk}
\eeq
which leave the $U(1)$ phases of $\zeta$ unaffected, 
$\ttau$ can be restricted to the torus $\re^n/(4 \pi \zahlen)^n$. 
In this gauge, the moduli describing the general caloron are the position vectors $\vec y_m$, 
comprised in $\vec y = (\vec y_1, \ldots, \vec y_n)$ and the torus coordinate $\ttau$ describing the $U(1)^{n-1}$ residual gauge symmetry and the temporal position of the caloron. 
Strictly speaking, these variables are coordinates on the cover of the moduli space of framed calorons. 
The true moduli space is obtained by dividing out the center of the gauge group.
This leads to orbifold singularities.

Under Fourier transformation, the Green's function $f_x$ (\refeq{df}) for calorons becomes 
$\hat f_x(z,z')\! \equiv\! \sum_{p,p'\in \zahlen} \!f_{x, p,p'} e^{2 \pi i(p
z - p' z')}$ 
and is a solution of the differential equation
\beq
\left\{\!\!\left(\!\frac{1}{2\pi i}\ddz \!-\! x_0\right)^{\!\!2} + \!\!\!\sum_{m \in \zahlen/n \zahlen} \!\!\!\chi_{[\mu_m, \mu_{m+1}]}(z) \, r_m^2 + \frac{1}{2 \pi} \!\!\!\sum_{m \in \zahlen/n \zahlen}\!\!\! \delta(z \!-\! \mu_m) |\vec y_m \!-\! \vec y_{m-1}| \!\right\}\!\!\hat f_x(z,z') = \delta(z-z'),\label{eq:dedf}
\eeq
in the gauge with $\hat A_0(z)$ constant.
Here $r_m = |\vec x - \vec r_m|$ is the center of mass radius of the $m^{\rm th}$ constituent. Expressions for $\hat f_x$ in other gauges are obtained by using that under the action of $\hat \cG$, $\hat f_x$ transforms as
\beq
\hat f_x(z,z') \rightarrow g(z) \hat f_x(z,z') g(z')^*, \quad g(z) \in \hat \cG . \label{eq:gaf}
\eeq
The Nahm construction of the $(1,1,\ldots,1)$ monopole, later obtained by as a special limit of the caloron, is discussed in the appendix.

\section{The caloron metric}
\subsection{Moduli spaces of \sd\ connections} \label{sec:modsp}
The metric on the moduli space $\cM$ of \sd\ connections on the manifold $M=\re^4/H$ is computed as the $L_2$ norm of its tangent vectors. 
These are gauge orthogonal variations of the connections with respect to their moduli. 
Specifically, $Z_\mu$ is tangent to the moduli space when it is a solution of the deformation equation and the gauge orthogonality condition requiring it to be a zero mode of the covariant derivative $D_\mu^{\rm ad}=\partial_\mu+[A_\mu,\cdot]$,
\beq
D^{\rm ad}_{[\mu}Z^{\vphantom{{\rm ad} }}_{\nu]} = \half \epsilon_{\mu\nu\alpha\beta}D^{\rm ad}_{[\alpha}Z^{\vphantom{{\rm ad} }}_{\beta]}, \quad D^{\rm ad}_\mu(A) Z^{\vphantom{{\rm ad} }}_\mu = 0.\label{eq:go}\label{eq:defeq}
\eeq
Written in terms of quaternions, these equations are concisely expressed as
$
D^{{\rm ad}\dagger} Z = 0,
$
from which one reads off the tangent space to admit three almost complex structures $I,J, K$ acting as $- \i, -\j, -\k$ on the right. Metric and \K\ forms read
\beq
(g, \hkom)_{\cal M}(Z, Z') = \frac{1}{4\pi^2}\int_{M} d_4 x \Tr\left(Z^\dagger(x) Z'(x) \right), \label{eq:mhkint}
\eeq
where $Z, Z'$ are any two tangent vectors. Gauge orthogonality of a general variation $\delta A_\mu$ of the \sd\ connection can be achieved by applying an infinitesimal gauge transformation $\Phi$,
\beq
Z_\mu = \delta A_\mu + D^{\rm ad}_\mu \Phi
,
\quad (D_\nu^{\rm ad})^2 \Phi = - D^{\rm ad}_\mu \delta A_\mu \label{eq:cohig}
.
\eeq
implying for the metric
\beq
g = \frac{-1}{4\pi^2}\int_M d_4 x \Tr ( \delta A_\mu- D^{\rm ad}_\mu (D_\nu^{\rm ad})^{-2}D^{\rm ad}_\rho \delta A_\rho)^2 . \label{eq:mhi}
\eeq

The \hk\ property of the moduli space follows formally from considering 
it as the infinite dimensional \hk\ quotient of the space of general connections $\cal A$
by the triholomorphic action of the group of gauge transformations $\cal G$\cite{AtiHit, DonKro}.
The moment map is $\mmu_\cG = \bar\eta_{\mu\nu} F_{\mu\nu} /8 \pi^2$, so that the zero set is formed by the space of self-dual solutions, which quotiented by $\cG$ gives the moduli space. That this quotient is well defined follows from the invariance of the \K\ forms 
\beq \hkom_{rs}\cdot \vec \sigma = \frac{1}{4\pi^2}\int_M d_4 x \bar\eta_{\mu\nu}
\Tr ( \delta_r A_\mu \delta_s A_\nu ),
\eeq
under infinitesimal gauge transformations, which is seen by adding arbitrary $D^{\rm ad}_\mu \Phi$ to the deformations. For the caloron the boundary condition \refeq{algau} is consistent with complex structures acting as $\bar\eta_{\mu\nu}$, \ie\ the non-trivial holonomy is compatible with the \hk\ structure.
One therefore expects caloron moduli spaces to be \hk .

For practical purposes the formal reasoning above is of little use.
Computing metrics on moduli spaces with the techniques presented depends crucially on the construction of the Green's function of the covariant Laplacian and 
in the present situation, we do not even have an expression for $A_\mu$ readily available. We take a different route which uses multi-instanton calculus, suitably adapted to the caloron situation. 
This allows for calculating the metric in terms of the ADHMN data and makes it thus feasible to find a compensating gauge transformation or to perform the \hk\ quotient. 

Moduli spaces of \sd\ connections can usually be written as a product of the
base space $M$, describing the center of mass and the non-trivial relative moduli space $\cM_\rel$,
\beq
\cM = M \times  \cM_\rel.
\eeq
In the metric this corresponds to a part describing the flat metric on the base space $M$ and one for the relative or centered metric on $\cM_\rel$, containing the nontrivial part. 
However, in the case at hand, where we want to take particular limits, it will be preferable to work with the full metric on $\cM$. 

\subsection{Isometric properties of the ADHM-Nahm construction.} \label{sec:isomadhmn}
We first recall the computation of the metric on the moduli space 
of instantons on $\re^4$ which can 
be entirely performed using ADHM techniques. 
Adapted to the caloron situation, 
this will translate into the formalism to 
calculate metrics in terms of Nahm data.

A tangent vector to the instanton \ms\ is given by 
\beq
Z_\mu(C) = v^\dagger(x) C \bar \sigma_\mu f_x u(x) \varphi^{-\half}(x) - 
\varphi^{-\half}(x) u^\dagger(x) f_x \sigma_\mu C^\dagger v(x), \label{eq:tv}
\eeq
where $C$ is a tangent vector to the \ms\ of ADHM data,
\beq
C = \left( \bea{c} c \\ Y \eea\right),
\eeq
which satisfies 
\beq
(\Delta^\dagger(x) C) = (\Delta^\dagger(x) C)^t. \label{eq:goc}
\eeq
Here the $\Re$ part is the deformation of the ADHM constraint and the $\Im$ part guarantees gauge orthogonality. 
Using an infinitesimal $U(k)$ transformation \refeq{symmadhm} $T = \exp(- i\delta X)$, where $\delta X = \delta X^\dagger$, the tangent vectors can be constructed as
\beq
C = \delta \Delta + \delta_X \Delta = 
\left( 
\bea{c} 
\delta \lambda + i\lambda \delta X\\
\delta B + i[B, \delta X]
\eea
\right), \label{eq:c}
\eeq
which automatically satisfy the deformation equation. Gauge orthogonality imposes
\beq
\tr\left(B^\dagger[B, i\delta X]- [B^\dagger, i\delta X] B + 2 i\delta X \Lambda + \lambda^\dagger \delta \lambda - \delta \lambda^\dagger \lambda + B^\dagger \delta B - \delta B^\dagger B \right) = 0. \label{eq:dxeq}
\eeq
The complex structures acting on tangent vectors $Z$ extend to $C$ in a natural way, $Z(C) \bar \sigma^i = Z(C\bar\sigma^i)$, as is seen from \refeq{tv} and $\sigma_\mu \bar \sigma^i = -\bar \eta_{\mu\nu}^i \sigma_\nu$. 
The metric can be evaluated using a powerful expression due to Corrigan \cite{CorUnp}, 
\beq
\Tr (Z^\dagger_\mu(x)Z'_\mu(x) ) = 
- \half \partial^2 \tr_2 \Tr \left(C^\dagger( 2 - \Delta(x) f_x \Delta^\dagger(x)) C' f_x\right), \label{eq:cor}
\eeq
The integral to compute the $L_2$ norm in \refeq{mhkint} is reduced to a boundary term corresponding with $x^2 \rightarrow \infty$, where $f_x$ is known, compare \refeq{fadhmas}. 
Using that $Z(C) \bar\sigma^i = Z(C \bar\sigma^i)$ and identifying the tangent space to the ADHM data with the vector space itself, the well-known (see also \cite{Mac}) hyperK\"ahler isometric property of the ADHM construction is proven
\beq
(g, \hkom)_\cM (Z,Z') = \Tr \left(Y^\dagger Y' + 2c^\dagger c' \right). \label{eq:adhmmet}
\eeq
The right hand side of \refeq{adhmmet} explains why \refeq{mkfadhm} gives the natural metric and \K\ forms on the space ${\hat{\cal A}}$ of ADHM matrices $\Delta$.
As the ADHM construction is an isometry and the moduli space of ADHM data $\mmu^{-1}(0)/ U(k)$ is \hk\ the same holds for the moduli space of instantons on $\re^4$. 

In employing the metric properties of the ADHM construction in the caloron case, one has -in addition to the deformation equation and gauge orthogonality- the algebraic gauge condition \refeq{algau} to be satisfied
\beq
Z_\mu(x+1) = \pl Z_\mu(x) {\cal P}_\infty^{-1}.\label{eq:agzm}
\eeq
This requires 
\beq
Y_{p,{p'}} = Y_{p-1, {p'}-1}, \quad c_{p+1} = {\cal P}_\infty c_p, \quad \delta X_{p,{p'}} =\delta X_{p-{p'}}.
\eeq
The compatibility of periodicity and nontrivial holonomy with the \hk\ structure on the level of the ADHM-Nahm construction can be seen from the complex structures acting on $Y$ and $c$ as multiplication by $-\i, -\j, -\k$ on the right. 

We define the Fourier transforms of the tangent vector
\beq
 \hat c(z)\! = \!\sum_{p\in \zahlen}\exp(-2\pi ipz)c_p = \sum_{m \in \zahlen/n \zahlen} \delta(z - \mu_m) \hat c_m, \quad \delta(z-z')\hat Y(z)\!=\!
\sum_{p,p'\in \zahlen}e^{2\pi i(pz-p' z')}Y_{p,p'},
\eeq
and find after Fourier transformation of eqs. (\ref{eq:goc}, \ref{eq:c}) the analogues of \refeq{defeq} as the deformation of the Nahm equation and a gauge orthogonality condition
\beqa
\ddz \hat Y_i(z) &=& -i\pi \sum_{m \in \zahlen/n \zahlen}\tr_2 \bar \sigma^i (\zeta^\dagger_m \hat c^{\vphantom \dagger}_m + \hat c^\dagger_m \zeta^{\vphantom \dagger}_m) \delta (z - \mu_m),\nonumber \\
\ddz \hat Y_0(z) &=& - i\pi \sum_{m \in \zahlen/n \zahlen} \tr_2 ( \zeta_m^\dagger \hat c_m - {\hat c}^\dagger_m \zeta_m)\delta (z - \mu_m) . \label{eq:ftgo}
\eeqa
To evaluate the caloron metric we use \refeq{cor} and closely follow the reasoning in \cite{NPB}. By Fourier transformation, Corrigan's formula is cast into
\beqa
\Tr Z^\dagger(x) Z'(x)\!\!\!& =&\!\!\! - \partial^2 \int_{S^1} dz \left( [ \hat Y^\dagger(z) \hat Y'(z) + {\hat c}^\dagger (z) < \hat c'> ] \hat f_x(z,z) \right) \\
\!\!\!&+&\!\!\! \half \partial^2 \int_{S^1} dz dz' \left([\cC(z) + \cY(z)] \hat f_x(z,z') [\cY_x'^{\dagger}(z') + \cC'^{\dagger}(z')] \hat f_x(z',z)\right),\nonumber \label{eq:ftcor}
\eeqa 
where we introduced the shorthand notation
\beq
\cC(z) = \hat c^\dagger(z) < \! \hat \lambda \! >, \quad \cY_x(z) = ( 2 \pi i)^{-1} \hat Y^\dagger(z) \hat D_x(z).
\eeq
In evaluating the integral over $\re^3\times S^1$, the $\partial_0^2$ term gives no contribution because of periodicity. The term involving $\partial_i^2$ is evaluated by partial integration as a boundary term at spatial infinity, for which the asymptotic behaviour of the Green's function $f_x(z,z')$ is needed. 
Since the asymptotic expression for the Green's function is independent of $n$,
\beq
\hat f_x(z,z') = \frac{\pi}{|\vec x|} e^{ - 2 \pi |\vec x| |z - z'| + 2 \pi i x_0 ( z - z')} + \cO({|\vec x|^{-2}}), \label{eq:fas}
\eeq
we can use the analysis for $SU(2)$ in \cite{NPB}. 
Combining the first line in \refeq{ftcor} with the only surviving term of the second, we find the following gauge independent expression
\beq
(g,\hkom)_{\cal M}(Z,Z') = 
\left(<\hat Y^\dagger Y' > + 2< \hat c^\dagger> < \hat c'> \right).
\label{eq:metev}
\eeq
This proves that the metric and \K\ forms on the caloron moduli space can be computed as the metric on the Nahm data. 
In other words, for $k=1$ $SU(n)$ calorons, the Nahm construction is a \hk\ isometry. 
A slightly modified proof shows this for monopoles of type $(1,1,\ldots,1)$ and can be found in the appendix. 

The isometric property is essential for what follows.
The metric on the caloron moduli spaces can now be calculated in terms of tangent vectors to the space of solutions to the Nahm equations,
with infinitesimal gauge transformations performed where needed.
This method, used in section \ref{sec:caldir}, is called direct as it concentrates on the gauge orthogonal tangent vectors to the moduli space.
An alternative method, given in section \ref{sec:hkqccm}, uses the fact that the moduli space of data is an infinite dimensional \hk\ quotient. 
It proceeds by using part of the $U(k)$ gauge symmetry to embed the moduli
in a finite dimensional \hk\ space. 
The metric on the moduli space is then found as the metric on a finite dimensional \hk\ quotient, with the remaining gauge action to be divided out. 

\subsection{Direct computation} \label{sec:caldir}
In the direct approach a compensating gauge function 
$
\delta \hat X(z)\! =\! \sum_{p\in \zahlen} X_p \exp( 2 \pi i pz)
$
has to be found to account for the tangent vectors
\beq
\hat c(z)= \sum_{m \in \zahlen/n \zahlen} \delta(z - \mu_m) \left(\delta \zeta_m +i \zeta_m \delta \hat X(\mu_m)\right), \quad 
\hat Y(z)= \frac{1}{2 \pi i} \left(\delta \hat A(z) + i\ddz \delta \hat X(z) \right)
\eeq
to be gauge orthogonal, \refeq{ftgo}. The gauge orthogonality of $\hat Y(z)$ implies for the compensating gauge function $\delta\hat X(z)$
\beq
\!- \frac{1}{2 \pi} \frac{d^2 \delta \hat X(z)}{dz^2} + 2 \delta \hat X(z) \!\!\!\!\!\sum_{m \in \zahlen/n \zahlen}\!\!\!\!\delta(z \!-\! \mu_m) |\vec\rho_m|\! = \!\!\!\!\!\sum_{m \in \zahlen/n \zahlen}\!\!\!\! \delta( z\! -\! \mu_m) \!\!
\left[ \frac{d \ttau_m}{4 \pi \nu_m} \!-\! \frac{d \ttau_{m-1}}{4 \pi \nu_{m-1}} - |\vec\rho_m|\vec\omw_m(\vec\rho_m) \!\cdot \! d \vec\rho_m \!\right]\!,\label{eq:dex}
\eeq
where we used \refeq{ident}. 
This differential equation implies that $\delta \hat X(z)$ is continuous and piecewise linear. 
Therefore, $\delta \hat X(z)$ is fully determined by the values $\delta \hat X_m$ it takes at $ z = \mu_m$, which are comprised in the vector $\delta\hat X = (\delta \hat X_1, \ldots, \delta\hat X_n)\in \re^n$. 
In the gauge chosen, all functions are either constants on the subintervals $(\mu_m, \mu_{m+1})$, or fixed by values at $z = \mu_m$. 
Therefore, the entire computation can conveniently be performed in terms of $n$ dimensional vectors and $n\times n $ matrix operators acting thereon, at the cost of introducing some extra notation. 
For taking derivatives, we will use the $n\times n$ matrix
\beq
S = \left(
\bea{ccccc}
1&\mmin 1&&&\\
&1&\mmin 1&&\\
&&\ddots&&\\
&&&1&\mmin 1\\
\mmin 1&&&&1
\eea
\right),
\eeq
with unspecified entries generally put to zero.
In addition we introduce the vector $ \vec\rho = (\vec\rho_1, \ldots, \vec\rho_n) \in \re^{3n}$ and diagonal matrices
\beq
N = \diag( \nu_1, \ldots, \nu_n), \quad
\vec\Omw = \frac{1}{4\pi}\diag(
\vec\omw_1(\vec\rho_1), \ldots, 
\vec\omw_n(\vec\rho_n)),
\quad 
V^{-1} =4 \pi \diag(\rho_1, \ldots, \rho_n).\label{eq:ddefs}\label{eq:mmcal}
\eeq
Inroducing the symbol $V$ anticipates its later interpretation as potential. 
In the sequel, all matrix multiplications between $n$-dimensional objects are implicitly assumed. The transpose $^t$ acts only on the indices running from $1$ to $n$. 

The Nahm connection is now represented by the $n$ dimensional vector
\beq
\hat A = (\hat A_1, \ldots, \hat A_n)^t = 2 \pi( N^{-1} \frac{\ttau}{4\pi} + \vec y \cdot \vec \sigma), 
\eeq
where $i \hat A_m$ is the value of $\hat A(z)$ on $(\mu_m,\mu_{m+1})$.
The Nahm equation reduces to
$\vec\rho = S^t \vec y$.
Similarly $\hat c(z) = \sum_{m \in \zahlen/n \zahlen} \delta(z - \mu_m) \hat c_m$ and $\hat Y(z) = i\sum_{m \in \zahlen/n \zahlen} \chi_{[\mu_m, \mu_{m+1}]} (z) \hat Y_m$ are fixed by
\beq
\hat c_m = \delta \zeta_m + i \zeta_m \delta \hat X_m, \quad
\hat Y = \frac{1}{2\pi}\delta \hat A -\frac{1}{2\pi} N^{-1} S \delta \hat X.
\label{eq:charc}
\eeq
Integrating the differential equation (\ref{eq:dex}) for $\delta \hat X(z)$ over small intervals $[\mu_m-\epsilon,\mu_m+\epsilon]$, $ \epsilon \downarrow 0$, gives conditions on the values $\delta \hat X_m$. This yields
\beq
\etp\left( S^t N^{-1} S + V^{-1} \right) \delta \hat X = ( S^t N^{-1} \frac{d\ttau}{4\pi} - V^{-1} \vec \Omw S^t \cdot d \vec y), \label{eq:charx}
\eeq
where we used that $ \int - d^2 \delta \hat X(z)/dz^2dz$ contributes
\beq
-\left( \delta \hat X'(\mu_m+) - \delta\hat X'(\mu_m-) \right) = -\left(
\frac{1}{\nu_m} \left( \delta \hat X_{m+1} - \delta \hat X_{m}\right)
-
\frac{1}{\nu_{m-1}} \left( \delta \hat X_{m} - \delta \hat X_{m-1}\right)\right).
\eeq
Eq. (\ref{eq:charx}) is solved by
\beq
\frac{\delta \hat X}{2\pi} =V S^t G^{-1}\frac{d \ttau}{4\pi} - 
 \left( 1 - V S^t G^{-1} S\right) \vec \Omw S^t\! \cdot\! d \vec y,
\eeq
such that
\beq
\hat Y
=
d \vec y \cdot \vec\sigma\!+\! N^{-1} \frac{d \ttau}{4\pi} \!-\! \etp N^{-1} S \delta\hat X=
 d \vec y \cdot \vec\sigma+ G^{-1}(\frac{ d \ttau}{4\pi} + S \vec \Omw S^t\! \cdot\! d \vec y)\nonumber,
\eeq
where we defined
$
G = N + S V S^t.
$
The integration over $S^1$ to evaluate the metric on the Nahm data in \refeq{metev} is carried out as 
$
< \hat Y ^\dagger \otimes\hat Y> = \hat Y ^\dagger N\otimes\hat Y
$
using that each subinterval has length $\nu_m = \mu_{m+1} - \mu_m$. 
Thus we obtain
\beqa
\half \tr_2 <\hat Y^\dagger \hat Y> \!\!\!&=&\!\!\! d \vec y\,^t \cdot N d \vec y + (\frac{d \tau}{4 \pi} + S \vec\Omw S^t \cdot d \vec y)^t G^{-1} N G^{-1} (\frac{d \tau}{4 \pi} + S \vec\Omw S^t \cdot d \vec y), \nonumber\\
\half <\! \hat Y^\dagger \!\wedge \!\hat Y \!>\!\!\! &=& \!\!\!- \half d \vec y\,^t N \wedge d \vec y\cdot \vec\sigma + \!\left( N G^{-1} (\frac{d \tau}{4 \pi} + S \vec \Omw S^t \cdot d \vec y) \right)^t\! \wedge d \vec y\cdot \vec\sigma. \nonumber
\eeqa
Using the properties (\ref{eq:ident},\ref{eq:psirmet}) of $\zeta_m$, 
the contribution to the metric of $\hat c_m$ defined in \refeq{charc} is found.
One obtains 
\beqa
 \tr_2 <\hat c^\dagger><\hat c> &=& d \vec y\,^t \cdot S V S^t d \vec y+ 
(\frac{ d \ttau}{4\pi} +S\vec \Omw S^t \cdot d \vec y)^t
 G^{-1}S V S^t G^{-1}(\frac{ d \ttau}{4\pi} 
+ S\vec \Omw S^t \cdot d \vec y),\nonumber\\
<\!\hat c^\dagger\!>\wedge<\!\hat c\!> &=& - \half ( S V S^t d \vec y)^t \wedge d \vec y\cdot \vec\sigma + \left( SVS^t G^{-1}(\frac{ d \ttau}{4\pi} + S \vec\Omw S^t \cdot d \vec y)\right)^t \wedge d \vec y\cdot \vec\sigma,\nonumber
\eeqa
where it is used that in the gauge chosen the phases of $\zeta$ are fixed.
The metric and \K\ forms on moduli space of the uncentered caloron are now readily obtained 
\beq
ds^2 = \ d \vec y\,^t G \cdot d\vec y + ( \frac{d \ttau}{4\pi} + \cW\cdot d \vec y)^t G^{-1} ( \frac{d \ttau}{4\pi} +\cW \cdot d \vec y),\label{eq:cm}
\eeq
\beq
\hkom = (\frac{ d \tau}{4 \pi} + \cW \cdot d \vec y)^t \wedge d \vec y - \half (G d \vec y)^t \wedge d \vec y,\label{eq:kfo}
\eeq
\[
G = N + S V S^t, \quad 
\cW = S \vec \Omw S^t \nonumber.
\]
Equivalently writing
\beq
G_{m,m'} = \nu_m \delta_{mm'} - \frac{ \delta_{m-1, m'}}{4\pi \rho_m} + \delta_{m,m'} \left( \frac{1}{4\pi \rho_m} + \frac{1}{4\pi\rho_{m+1}} \right)- \frac{\delta_{m+1, m'}}{4\pi \rho_{m+1}}, \quad m, m' \in \zahlen/n \zahlen,
\eeq
reveals the form of $G$ as given in \cite{LeeYi1}; thus we confirm the conjectured form for the metric in \cite{LeeYi1}. 
As is readily checked, from eqs. (\ref{eq:dirac}, \ref{eq:ddefs}) it follows that $G$ and $\cW$ satisfy the hyperK\"ahler conditions (\ref{eq:hkcon})
\beq
\vec\nabla_y G = \vec \nabla_y \times \cW, \quad
\partial^i_m G_{m',m''} = \epsilon_{ijk} \partial^j_m \left(\cW \right)^k_{m',m''}, 
\eeq
($ \partial^i_m = \partial/\partial y^i_m$), which 
implies the \K\ forms in (\ref{eq:kfo}) to be closed and the caloron metric to be \hk . 

The metric has $n$ commuting triholomorphic isometries,
\beq
	\frac{\partial}{\partial\ttau_m}, \quad m = 1, \ldots, n,
\eeq
as $G$ and $\cW$ are $\ttau$ independent. The isometries correspond to shifts on the $n$-torus $\re^n/(4 \pi \zahlen)^n$ which describe the residual $U(1)^{n-1}$ gauge invariance and the temporal position
\beq
\xi_0 = \frac{1}{4 \pi} \sum_{m \in \zahlen/n \zahlen} \ttau_m \in S^1,
\eeq 
of the caloron. Therefore, the caloron moduli space is a toric \hk\ manifold, with dimension $4n$. $3n$ coordinates describe the monopole positions and $n$ phase angles parameterise the temporal position and
residual $U(1)^{n-1}$ gauge invariance in the case of maximal symmetry breaking.
From the uncentered caloron metric in \refeq{cm}, all other metrics discussed in this paper can be obtained by taking suitable limits. 
In the next subsection the caloron metric will be obtained using the \hk\ quotient. 

The non-trivial part of the metric is obtained by splitting off the center of mass coordinate $\xi$ in \refeq{com}. 
To this aim, we express the metric in terms of $\xi$ and $n-1$ relative monopole position vectors $\vec\rho_m$, using that $\vec\rho _n = - \sum_{m=1}^{n-1} { \vec\rho_m}$ because of \refeq{conrho}.
The two sets of coordinates are related by the $n \times n$ dimensional "centering matrix" $F_\ca$,
\beq
F_\ca = (S_c, N \e), \quad \left(\bea{c}\tilde{\vec\rho}\\\vec\xi\eea\right) =
 F^t_\ca \vec y. 
\eeq
Here, the $n \times (n-1)$ dimensional matrix $S_c$ is obtained from $S$ by omitting its last column, and we defined $\e = (1, \ldots, 1)^t \in \re^{n}$.
A tilde denotes from now on the restriction to the first $n-1$ coordinates, \eg\ $\tilde{\vec \rho} = (\vec \rho_1, \ldots, \vec \rho_{n-1})^t$. 
New torus coordinates $\tilde \upsilon = (\upsilon_1, \ldots, \upsilon_{n-1})^t$ are introduced as well
\beq 
\tau = F_\ca \left( \bea{c} \tilde\upsilon\\ 4 \pi \xi_0\eea\right). \label{eq:tidy}
\eeq
The centered metric will be again \hk, as splitting of the center off mass metric 
amounts to taking the \hk\ quotient under the $U(1)$ action
\beq
\tau_m \rightarrow \tau_m + \nu_m t_c, \quad m = 1, \ldots, n, \quad t_c \in \re.
\eeq
From eqs. (\ref{eq:cm}, \ref{eq:kfo}) it is seen that this action is a triholomorphic isometry whose moment map gives the center of mass of the caloron
\beq
\mmu = \frac{1}{4\pi}\sum_{m\in \zahlen/n \zahlen} \nu_m \vec y_m = \frac{\vec\xi}{4 \pi}.
\eeq
Indeed, the phase variables $\tilde\upsilon$ are invariant under the $U(1)$ action and can serve as coordinates on the quotient whereas the fibre coordinate $\xi_0$ changes as $\xi_0\rightarrow\xi_0+t_c$.

In the new basis the relative metric is expressed in terms of a relative mass matrix and relative interaction potentials
\[
F^{-1}_\ca G(F^{-1}_\ca)^t \! = \!\left(\!\!
\bea{cc}
\tilde G_\rel&\\
&1
\eea
\!\!\right), \quad \tilde G_\rel \!=\! \tilde M + \tilde V_\rel
, \]
\beq
(\tilde V_\rel)_{mm'}\! =\!
\tilde V_{mm'}+\frac{1}{4 \pi |\vec\rho_n|}, \quad
(\tilde \cW_\rel)_{mm'} \! = \! 
\tilde \cW _{mm'}+\frac{ \vec \w_n( \vec\rho _n)}{4 \pi} .\label{eq:gmv}
\eeq
where $m,m' = 1,\ldots, n-1$, $\vec\rho _n = - \sum_{m=1}^{n-1} { \vec\rho_m}$.
The relative mass matrix $\tilde M$ is defined as 
\beq
F_\ca^t N^{-1} F_\ca \!=\! \left(\!\bea{cc} \tilde M^{-1}&\\&1\eea\!\right), \quad
\tilde M^{-1}\!=\!\left(\!\!\!
\bea{cccccc}
\frac{1}{\nu_n}+\frac{1}{\nu_1}&-\frac{1}{\nu_1}&&&&\\
-\frac{1}{\nu_1}&\frac{1}{\nu_1}+\frac{1}{\nu_2}&-\frac{1}{\nu_2}&&&\\
\ddots&\ddots& \ddots&&\\
&-\frac{1}{\nu_{n-3}}&\frac{1}{\nu_{n-3}}+\frac{1}{\nu_{n-2}}&-\frac{1}{\nu_{n-2}}\\
&&-\frac{1}{\nu_{n-2}}&\frac{1}{\nu_{n-2}}+\frac{1}{\nu_{n-1}}
\eea
\!\!\!\!\!\!\!\!\right),
\eeq
its explicit form allowing one to take limits that correspond to massless monopoles
\beq
\tilde M = \tilde M^t, \quad \tilde M_{mm'} = 
(\nu_m+ \cdots +\nu_{n-1})(1-\nu_{m'}\cdots-\nu_{n-1}) \quad {\rm for} \, m \ge m', \quad m,m' = 1, \ldots, n-1.\label{eq:relmmcal}
\eeq
The centered metric and \K\ forms now read
\[
g = d \xi_\mu d \xi_\mu + d \tilde { \vec\rho }\,^t \tilde G_\rel \cdot d\tilde{ \vec\rho } + ( \frac{d \tilde\upsilon}{4 \pi} + \tilde \cW_\rel \cdot d \tilde{ \vec\rho })^t \tilde G^{-1}_\rel ( \frac{d \tilde\upsilon}{4 \pi} + \tilde \cW_\rel \cdot d \tilde{ \vec\rho }),\]
\beq
\hkom = d \xi_0 \wedge d \vec \xi - \half d \vec\xi \wedge d \vec \xi
+ ( \frac{d \tilde\upsilon}{4 \pi} + \tilde \cW_\rel \cdot d \tilde{ \vec\rho })^t
\wedge d \tilde{ \vec\rho } - \half ( \tilde G_\rel d \tilde{ \vec\rho })^t \wedge d \tilde { \vec\rho }. \label{eq:mcc}
\eeq
The first terms give the center of mass metric on $\re^3 \times S^1$, the other terms represent the non-trivial part of the metric. Both are toric \hk, and have an $SO(3)$ invariance corresponding to spatial rotations. 

\subsection{HyperK\"ahler quotient construction}\label{sec:hkqccm}
We follow the approach in \cite{MurNot} for BPS monopoles of type $(1,1, \ldots,1)$ and consider the right hand side of \refeq{metev} as the natural metric on the space of caloron Nahm data ${\hat{\cal A}}$ 
\beq
(g,\hkom)_{\hat{\cal A}} = 
\left(<d\hat A^\dagger \otimes d\hat A> + 2< d\hat \lambda^\dagger> \otimes< d\hat \lambda> \right).
\eeq
One then notes that the group $\hat \cG$ of $U(1)$ gauge transformations on $\hat S^1$ acts triholomorphically on ${\hat{\cal A}}$. 
The zero set of the associated moment map is formed by the set $\cN$ of solutions to the Nahm equations, which after quotienting by the $U(1)$ gauge action ${\hat {\cal G}}$ on the dual $S^1$ gives the moduli space of Nahm data. By virtue of \refeq{metev} this quotient is isometric to the caloron moduli space, 
\beq
{\cal M} = {\cal N}/{\hat {\cal G}}.
\eeq
As both $\cal N$ and $\hat {\cal G}$ are infinite dimensional, it is not obvious that this procedure is well-defined. 
However, using the gauge action we can restrict to those solutions $\cN_0$ to the Nahm equations which have constant $\hat A^0(z)$ on the subintervals $(\mu_m, \mu_{m+1})$. 
As the Nahm equations force $\hat A_i(z)$ to be piecewise constant, there are 
$n$ quaternions specifying the Nahm connection, denoted by $y \in \qu^n$. 
The singularities (or matching data) are
described by $n$ complex two-component vectors $\zeta_m$, denoted by $\zeta\in \complex^{n,2}$. Hence, ${\cal N}_0$ is a subset of the space ${{{\hat{\cal A}}_0}} = \qu^n \times \complex^{n,2}$ of possible piecewise constant data,
which has metric and K\"ahler forms
\beq
(g, \hkom) = d y^\dagger \otimes N d y + 2 d \zeta^\dagger \otimes d \zeta \label{eq:kmfv},
\eeq
as is natural from \refeq{metev}. 
On ${\hat{\cal A}}_0$, the gauge action $\hat {\cal G}$ is restricted 
to the set $\hat \cG_0$ of gauge functions with piecewise linear and continuous $\log$. 
These are determined by the values $h$ assumes at $z = \mu_m$. 
Under these gauge transformations, 
$\hat A$ and $\zeta$ change according to
\beq
\zeta_m \rightarrow e^{i t_m} \zeta_m,
\quad \psi \rightarrow \psi + 2 t, 
\quad y \rightarrow y - \frac{1}{2\pi}N^{-1} S t
\label{eq:act},
\eeq
where $t = (h(\mu_1), \ldots, h(\mu_n))\in \re^n/(2 \pi \zahlen)^n$ and $\psi= (\psi_1, \ldots, \psi_n)/(4 \pi \zahlen)^n$ denotes the phases of $\zeta$. The lattices correspond to gauge transformations of type (\ref{eq:intijk}). 
Therefore the action of the restriction $\hat\cG_0$ 
of $\hat\cG$ on ${\hat{\cal A}}_0$ is equivalent 
to an $\re^n$ action on $\qu^n \times \complex^{n,2}$.
Thus we reduced the infinite dimensional \hk\ quotient to a finite dimensional.
This technique was also used for the $\oo$ monopole metric \cite{MurNot}.
The metric on the moduli space of Nahm data can now be computed as a metric on a \hk\ quotient of a finite dimensional euclidean space by a toric group action.
To do this we follow \cite{GibRycGot}.
From the metric and \K\ forms on ${\hat{\cal A}}_0$, 
determined by inserting 
eqs. (\ref{eq:defs}, \ref{eq:psirmet}) in \refeq{kmfv},
\beqa
&&ds^2 = dy^\dagger N d y + d \vec\rho\,^t V \cdot d \vec\rho + ( \frac{d \psi}{4\pi} + \vec\Omw \cdot d \vec\rho)^t V^{-1} ( \frac{d \psi}{4\pi} + \vec\Omw \cdot d \vec\rho ).\\
&&\hkom = -\half (N d \vec y)^t \wedge d \vec y + ( N dy_0)^t \wedge d \vec y + (\frac {d \psi}{4 \pi} + \vec\Omw \cdot d \vec\rho )^t \wedge d \vec\rho -\half ( V d \vec\rho )^t \wedge d \vec\rho,\nonumber
\eeqa
the action (\ref{eq:act}) is seen to be triholomorphic. 
The moment map for this $\re^n$ action reads
\beq
\mmu \cdot \vec\sigma= - \frac{1}{4\pi} S^t(y - \bar y) - \myIm i \zeta^\dagger P \zeta,
\eeq
where $P = ( P_1, \ldots, P_n)^t$, 
and has a zero set $\mmu^{-1}(0)$ given by the solutions $\hat A$ corresponding to $\vec\rho = S^t \vec y$.
Therefore, the space of piecewise constant solutions to the Nahm data is $(\hat A, \zeta) \in\cN_0 =\mmu^{-1}(0) \subset {\hat{\cal A}}_0$. The moduli space of Nahm data is this set quotiented by the reduction of the gauge action in \refeq{ga}, or equivalently $\re^n$. Hence 
\beq
\cM = \cN/\hat\cG = \cN_0/\hat\cG_0 = \mmu^{-1}(0)/\re^n.
\eeq 
The metric on $\mmu^{-1}(0)$ reads
\beqa
&&ds^2 = d \vec y( S V S^t + N) d \vec y 
+ (\frac{ d \psi}{4\pi} + \vec\Omw \cdot S^t d \vec y)^t V^{-1} ( \frac{d \psi}{4\pi} + \vec\Omw \cdot S^t d\vec y) + d y_0^t N d y_0\\
&&\hkom = (\frac{ d \tau}{4 \pi} + \cW \cdot d \vec y)^t \wedge d \vec y - \half (G d \vec y)^t \wedge d \vec y. \label{eq:omopm}
\eeqa
The $n$ vector 
\beq
\frac{\tau}{4\pi} = S \frac{\psi}{4\pi} + N y_0,
\eeq
is invariant under the $\re^n/(2 \pi \zahlen)^n$ action (\ref{eq:act}) and can therefore be used as coordinate on the quotient $\mmu^{-1}(0)/\re^n=\cM$, together with $\vec y$. Cotangent vectors involving $d \psi$ have a vertical component, \ie\ lie along the $\re^n$ fibre. 
The horizontal and vertical part of the metric 
are separated by inserting 
$ y_0 = \frac{1}{4\pi} N^{-1} (\tau -S \psi)$ 
and completing the squares to obtain
\beqa
ds^2 \!\!&=&\!\! 
d \vec y\,^t G d \vec y +\frac{ d \tau^t}{4\pi} N^{-1} \frac{d \tau}{4\pi}+ d \vec y\,^t S \cdot \vec \Omw V^{-1} \vec\Omw \cdot S^t d \vec y \nonumber\\&&\quad - ( S^t N^{-1} \frac{d \tau}{4\pi} - V^{-1} \vec \Omw S^t \cdot d \vec y)^t
\frac{1}{V^{-1} + S^t N^{-1} S}( S^t N^{-1}\frac{ d \tau}{4\pi} - V^{-1} \vec \Omw S^t \cdot d \vec y) \nonumber\\
&&\quad
+\varphi^t ( {V^{-1} + S^t N^{-1} S}) \varphi, \label{eq:mbhp}
\eeqa
where the one form $\varphi$ denotes the component along the $\re^n$ fibre
\beq
\varphi = \frac{d \psi}{4\pi} + \frac{1}{V^{-1} + S^t N^{-1} S}V^{-1} \vec\Omw S^t \cdot d \vec y - \frac{1}{V^{-1}+ S^t N^{-1} S}S^t N^{-1} \frac{d \tau}{4\pi}. 
\eeq
Horizontal projecting to the metric on $\mmu^{-1}(0)/\re^n$ amounts to discarding the last term in \refeq{mbhp} and one obtains (after reorganising) the metric on the caloron moduli space $\cM$ given in \refeq{cm}. 
For the \K\ forms, this projection is generally not necessary: \refeq{omopm} is precisely the \K\ form in \refeq{kfo}. 
This is a manifestation of the degeneracy of the \K\ forms along the gauge orbit, needed for the \hk\ quotient to be well defined.

\section{Instanton and monopole limits of the caloron}
From the caloron metric, other toric \hk\ manifolds can be
obtained by taking suitable limits.
For large $\cT$ or equivalently all 
$\rho_m$ small, 
one expects the metric to approach 
the moduli space for $k=1$ $\sun$ instantons on $\re^4$.
To study this limit, we consider the centered metric \refeq{mcc}.
For small $\rho_m$, the 
elements of the relative mass matrix $\tilde M$ in \refeq{gmv} 
are dominated by the $\rho_m^{-1}$ terms in $\tilde V_\rel$,
\beq
F^{-1}_\ca G (F^{-1}_\ca)^ t = \left( \bea{cc} \tilde G_\rel &\\&1 \eea \right) \rightarrow 
\left ( \bea{cc} \tilde V_\rel&\\&1 \eea\right), \quad \rho_m \rightarrow 0, \quad m = 1, \ldots, n-1,
\eeq
resulting in the asymptotic form for the non-trivial part of the metric and \K\ forms
\[
g_{\rm limit} = 
d \tilde { \vec\rho }\,^t \tilde V_\rel \cdot d\tilde{ \vec\rho } + ( \frac{d \tilde\upsilon}{4 \pi} + \tilde \cW_\rel \cdot d \tilde{ \vec\rho })^t \tilde V^{-1}_\rel ( \frac{d \tilde\upsilon}{4 \pi} + \tilde \cW_\rel \cdot d \tilde{ \vec\rho }),\]
\beq
\hkom_{\rm limit}= 
( \frac{d \tilde\upsilon}{4 \pi} + \tilde \cW_\rel \cdot d \tilde{ \vec\rho })^t
\wedge d \tilde{ \vec\rho } - \half ( \tilde V_\rel d \tilde{ \vec\rho })^t \wedge d \tilde { \vec\rho }.\label{eq:glim}
\eeq
The caloron with trivial gauge holonomy has the same limiting metric,
as follows directly from taking the limit $\nu_1, \ldots, \nu_{n-1} \rightarrow 0, \nu_n \rightarrow 1$ of the caloron relative mass matrix in \refeq{relmmcal}.
The phase variables are now given by $\upsilon_m =
\tau_m + \ldots + \tau_{n-1} \in \re/(4 \pi \zahlen)$, cf. \refeq{tidy}.
The \K\ forms $\hkom_{\rm limit}$ are closed, since the \hk\ conditions (\ref{eq:hkcon}) are satisfied
\beq
\vec\nabla_\rho \tilde G_\rel = \vec \nabla_\rho \times \tilde \cW_\rel,
\eeq
hence the limiting metric for large $\cT$ is \hk. It is known as the Calabi metric. 

This limit was discussed in \cite{LeeYi1} using indirect arguments. 
With the techniques presented in this paper, it is easy to
prove explicitly that the limiting metric is indeed the metric for both the ordinary $k=1$ $\sun$ instantons on $\re^4$ and the calorons with trivial holonomy. 
It follows immediately when realising that the $4(n-1)$ dimensional Calabi space can be obtained as the \hk\ quotient 
of $\qu^{n}$ by a $U(1)$ action \cite{GibRycGot}. 
This quotient emerges naturally 
from both the construction of the charge one $\sun$ instanton and the trivial holonomy caloron. 
First note that there is a one to one correspondence between the ADHM data of the $k=1$ $SU(n)$ instanton and the Nahm data of the trivial holonomy caloron in the $\hat \cG$ gauge with constant $\hat A_0(z)$. 
The latter are given in terms of $(\xi, \zeta) \in \qu \times \complex^{n,2}$ as $\hat A(z) = 2 \pi i \xi,\quad \hat \lambda(z) = \delta(z) \zeta$ and directly translate into ADHM data $\lambda = \zeta, \quad B = \xi$ for the instanton. 
With only one subinterval, the metric on the Nahm data now reduces to the expression for the instanton (\ref{eq:mkfadhm}). 
Having restricted to constant $\hat A_0(z)$, the remaining transformations in $\hat {\cal G}_0$ leave $\xi$ invariant, apart from confining $\xi_0$ to the circle through $g(z) = \exp( 2 \pi i p z), p \in \zahlen$. 
For their action on the matching data only the $U(1)$ formed by the values $g(0)$ is relevant. 
Therefore, in both cases the nontrivial part of the moduli space is the quotient
of $\complex^{n,2}$ (with $(g, \hkom) = 2d \zeta^\dagger \otimes d \zeta$) by the $U(1)$ action 
\beq
\zeta_m \rightarrow e^{it} \zeta_m,\quad \psi_m \rightarrow \psi_m + 2 t, \quad \quad m=1, \ldots n, \quad t \in \re/(2 \pi \zahlen). \label{eq:ueen}
\eeq 
(Identifying $\complex^2$ and $\qu$, this quotient is readily seen to be equivalent to that discussed in eq. (36) of \cite{GibRycGot}).
The corresponding moment map, zero set and invariants are given by
\beq
\mmu = \frac{1}{2 \pi}\sum_{m \in \zahlen/n \zahlen} \vec\rho _m, \quad
\sum_{m \in \zahlen/n \zahlen} \vec\rho = 0, \quad 
\tilde \upsilon_m = \psi_m - \psi_n, \quad m = 1, \ldots, n -1.
\eeq
Expressing the metric on the zero set in terms of invariants and the terms involving $d \psi_n$ describing the fibre part, one obtains \cite{GibRycGot}
the Calabi metric in \refeq{glim}. 

The Calabi metric has an $SU(n)$ triholomorphic isometry, reflecting the $\sun$ 
gauge symmetry of the $k=1$ instanton and trivial holonomy caloron.
As explained in section \ref{sec:prel} for the instanton, it emerges
as the $SU(n)$ acting on $\zeta$ in \refeq{ueen} on the left, commuting with $U(1)$, and descending to the quotient.
A direct calculation using a compensating gauge transformation gives the same result.

In \cite{NPB}, \cite{LAT}, it was explicitly shown from the action density that removing one of the constituent monopoles of the caloron to spatial infinity, $|\vec y_n| \rightarrow \infty$ turns it into a static \sd\ $\sun$ solution, \ie\ a monopole in the BPS limit. Indeed, this limit corresponds to the compactifaction length going to zero.
The Nahm data suggest that the remnant is the $(1,1,\ldots,1)$ monopole. 
We will show indeed that the metric in this limit has the required form. 

Removing a constituent is described by a \hk\ quotient. 
Consider the $U(1)$ action that changes the phase of the $m$th monopole in the uncentered caloron
\beq
\tau_m \rightarrow \tau_m + t, \quad t \in \re/(4 \pi\zahlen).
\eeq
It is a triholomorphic isometry as follows from eqs. (\ref{eq:cm},\ref{eq:kfo}).
Its moment map $\mmu_{\rm fix}$ is exactly the position of the $m$th monopole, $\mmu_{\rm fix} = \vec y_m/(4 \pi)$. 
Therefore, the metric on the quotient, the caloron moduli space with the $m$th constituent fixed, is \hk\ irrespective of its position. 
For finite $|\vec y_m|$, the resulting metric on the quotient $\mmu^{-1}_{\rm fix}(\vec y_m)/\re$ is complicated, and no longer $SO(3)$ symmetric. 
Removing the constituent, $|\vec y_m| \rightarrow \infty$, \ie\ fixing it at spatial infinity, gives the \hk\ metric of
the remnant BPS monopole, with a simple form and $SO(3)$ symmetry restored.

The metric with the $n$th monopole far away, in which case $\rho _1^{-1}, \rho_{n}^{-1}\rightarrow 0$, reads
\beq
(g, \hkom) = (g_n, \hkom_n) + (g_\mon, \hkom_\mon).
\eeq
Here the removed monopole is described by
$
g_n = \nu_n d \vec y_n^2 + {\nu_n}^{-1} d\tau_n^2, 
$
and the remnant by
\[
g_\mon = d \vec y_{\rm m}\,^t G_\mon d \vec y_{\rm m} + ( \frac{d\tau_\mon}{4 \pi} + \cW_\mon \cdot d \vec y_\mon)^t G^{-1}_\mon ( \frac{d\tau_\mon}{4 \pi} + \cW_\mon \cdot d \vec y_\mon),
\]
\beq
\hkom_\mon=-\half ( G_\mon d \vec y_\mon)^t \wedge d \vec y_\mon + ( \frac{d \tau_\mon}{4 \pi} + \cW_\mon \cdot d \vec y_\mon)^t \wedge d \vec y_\mon, \label{eq:lwyu}
\eeq
where 
\[ 
G_{\rm m} = N_{\rm m} + S_{\rm m} V_{\rm m} S_{\rm m}^t, \quad \cW_{\rm m} = S_{\rm m} \vec W_{\rm m} S_{\rm m}^t,
\]
\[
 V_{\rm m}^{-1} =4 \pi \diag( \rho_2, \ldots,\rho _{n-1}), \quad
\vec W_{\rm m} = \diag( \vec \w_2(\vec\rho_2), \ldots, \vec \w_{n-1} ( \vec\rho _{n-1}))/(4 \pi),
\]
\beq
N_{\rm m} = \diag (\nu_1, \ldots, \nu_{n-1}), \quad\!\!\! y_{\rm m} = ( y_1, \ldots,y_{n-1})^t, \quad\!\!\!\vec\rho_{\rm m} = ( \vec\rho_2, \ldots, \vec\rho _{n-1})^t,\quad \!\!\!\tau_\mon = (\tau_1, \ldots, \tau_{n-1})^t, \label{eq:mmmon}
\eeq
\beq
S_{\rm m} = \left( \bea{cccc}
-1 &&&\\
1&-1&&\\
&\ddots&\ddots&\\
&&1&-1\\
&&&1
\eea
\right)\in \re^{n-1, n-2}. \label{eq:sm}
\eeq
More explicitly, the potential term in \refeq{lwyu} reads
\beq
4\pi S_\mon V_\mon S_\mon^t=\left(
\bea{ccccc}
\frac{1}{\rho_2}&-\frac{1}{\rho_2}&&&\\
-\frac{1}{\rho_2}&\frac{1}{\rho_2}+\frac{1}{\rho_3}&-\frac{1}{\rho_3}&&\\
\ddots&\ddots& \ddots\\
&-\frac{1}{\rho_{n-2}}&\frac{1}{\rho_{n-2}}+\frac{1}{\rho_{n-1}}&-\frac{1}{\rho_{n-1}}\\
&&-\frac{1}{\rho_{n-1}}&\frac{1}{\rho_{n-1}} 
\eea
\right) \in \re^{n-1, n-1}
\eeq
The vector potential $\cW_\mon$ has a similar structure.
The metric in \refeq{lwyu} is that of the uncentered $SU(n)$ monopole of type $(1,1, \ldots,1)$. 
The calculation of the metric on its space of Nahm data was performed in \cite{MurNot,GibRycGot}. 
Details on the Nahm construction of the $\oo$ monopole and a proof of its isometric property as well as an outline of the calculation of the metric can be found in the appendix. 

To connect with \cite{LeeWeiYi1}, we have to center the monopole. We introduce
\beq
F_\mon = \left(\bea{cc} S_\mon, &\frac{1}{\nu}N_\mon \e_\mon \eea \right)\in \re^{n-1, n-1}
\eeq
where $\e_\mon = (1, \ldots, 1) \in \re^{n-1}$ and $\nu = \sum_{m=1}^{n-1} \nu_m $ denotes the mass of the monopole. 
The relative position variables $\vec\rho_\mon$ are reinstated
and the center of mass $\re^3$ position is separated off using
\beq
\vec y_\mon=(F_\mon^t)^{-1}\left( \bea{c}\vec\rho_\mon\\\vec\xi_\mon \eea\right), \quad 
\vec\xi_\mon = \frac{1}{\nu} \sum_{m=1}^{n-1} \nu_m \vec y_m.
\eeq
The mass matrix in this basis is given by
\[
F_\mon^t N_\mon^{-1} F_\mon = \left(\bea{cc} M_\mon^{-1}&\\& \nu^{-1}\eea\right), \quad 
M_\mon^{-1}=\left(\!
\bea{ccccc}
\frac{1}{\nu_1}+\frac{1}{\nu_2}&-\frac{1}{\nu_2}&&&\\
-\frac{1}{\nu_2}&\frac{1}{\nu_2}+\frac{1}{\nu_3}&-\frac{1}{\nu_3}&&\\
\ddots&\ddots& \ddots&&\\
&-\frac{1}{\nu_{n-3}}&\frac{1}{\nu_{n-1}}+\frac{1}{\nu_{n-2}}&-\frac{1}{\nu_{n-2}}\\
&&-\frac{1}{\nu_{n-2}}&\frac{1}{\nu_{n-2}}+\frac{1}{\nu_{n-1}}
\eea
\!\!\!\!\!\!\right),
\]
\beq
M_\mon = M_\mon^t, \quad (M_{\mon})_{ m,m'} = \nu^{-1}(\nu_1 + \cdots + \nu_{m})(\nu_{m'+1} + \cdots +\nu_{n-1}), \quad {\rm for}\, m' \ge m. \label{eq:relmmmon}
\eeq
Furthermore, alternative torus coordinates $\chi_\mon = (\chi_1, \ldots, \chi_{n-2})$ are introduced, as well as a global $U(1)$ phase $\xi_{0,\mon}$
\beq
 \tau_\mon =F_\mon \left(\bea{c} \chi_\mon\\\xi_0 \eea\right), \quad
\xi_{0, \mon} = \sum_{m=1}^{n-1} \tau_m.
\eeq
In the new coordinates, the uncentered metric is the sum of the center of mass and relative metric 
\beq
g_\mon = \nu d \vec\xi_\mon \cdot d\vec \xi_\mon + \nu^{-1} d \xi_{0, \mon}^2 + g_\mon^c,
\eeq
where the nontrivial part \beq
g_\mon^c = d \vec\rho _\mon^t (M_\mon + V_\mon) \cdot d \vec\rho_\mon + 
(\frac{d \chi_\mon}{4\pi} + \vec\Omw_\mon \cdot d \vec\rho_\mon)^t (M_\mon+V_\mon)^{-1} (\frac{d \chi_\mon}{4\pi} + \vec\Omw_\mon \cdot d \vec\rho_\mon) \label{eq:lwy}
\eeq
is the Lee-Weinberg-Yi metric \cite{LeeWeiYi1}. It is of \thk\ form. %
Thus we proved that the $(1,1,\ldots,1)$ monopole is a limit of the caloron, identifying the static remnant in \cite{MC}\cite{LAT}.

Finally, we note that the $\oo$ monopole has only one magnetic winding, as explained in the introduction. It is opposite to the winding of the removed monopole, and hence, we can apply the reasoning in \cite{TauTop} explaining how the instanton charge arises also for $SU(n)$ from braiding two monopoles \cite{NPB}. 

\section{Discussion}
Since the metric describes the Lagrangian for adiabatic motion on the moduli space \cite{ManRem}, it reflects the interactions of the monopole constituents. 
The constituent nature of the caloron solution, easily extracted from the action density, should therefore also be reflected in the metric. 
The action density of the $k=1$ $SU(n)$ caloron \cite{MC} is derived from \refeq{ad} employing Green's function techniques and reads
\beq
- \half {\rm Tr} F_{\mu\nu}^2 = - \half \partial_\mu^2 \partial_\nu^2 \log
\Psi.
\label{eq:adpsi}
\eeq
Here the positive scalar potential $\Psi$ is defined as 
\beq
\Psi(x)=
\half\tr\prod_{m=1}^n\left\{ A_m\right\}
-\cos(2\pi x_0),
\label{eq:calad}
\eeq
where 
\beq
A_m = \left(\bea{cc}r_m&|\vec y_m-
\vec y_{m+1}|\\0&r_{m+1}\eea\right)\left(\bea{cc}
c_m&s_m\\s_m&c_m\eea
\right)\frac{1}{r_m}\eeq
given in terms of the center of mass radii $r_m=|\vec x-\vec y_m|$ of the
$m^{\rm th}$ constituent monopole, $c_m =\cosh(2\pi\nu_mr_m) $, $s_m = \sinh(2\pi\nu_mr_m)$ and $\prod_{m=1}^n A_m = A_n \cdots A_1$.
The energy density for the $(1,1,\ldots,1)$ monopole is obtained from it by sending the $n$th constituent to infinity, which gives \cite{LAT}
\beq
{\cal E}(\vec x)=-\half\tr F_{\mu\nu}^{\,2}(\vec x)=-\half\Delta^2
\log\tilde\Psi_\mon(\vec x),\label{eq:moned}
\eeq
\beq
\tilde\Psi_\mon(\vec x)=\half\tr\left\{\frac{1}{r_{n-1}}
\left(\bea{cc}
s_{n-1}&c_{n-1}\\0&0\eea\right)\prod_{m=1}^{n-2}
A_m\right\}.
\eeq
(see \cite{Lu} for some special cases). 
These densities allow for an unambiguous identification of elementary BPS monopoles as constituents of calorons, and $(1,1,\ldots,1)$ monopoles, as in the limit where $r_m\ll r_l$ for all $l\neq m$ the action density approaches that of the single BPS monopole \cite{MC}. 
The corresponding limit in the uncentered metrics reveals
\beq
ds^2|_m = \nu_m d \vec y_m \cdot d \vec y_m + \frac{1}{\nu_m} d\tau_m^2
\label{eq:r3m} 
\eeq
for the part describing the $m^{\rm th}$ constituent, as all interaction potentials approach zero with the other constituents far away.
Eq. (\ref{eq:r3m}) is the flat metric on $\re^3 \times S^1$, the twofold cover of the moduli space for the elementary BPS monopole.
Therefore the limit of the moduli space corresponding to all monopoles well separated- of the (cover of the) caloron moduli space
can be seen as a product of elementary BPS monopole moduli spaces. 

We obtained the metric for the $k=1$ $SU(n)$ caloron assuming symmetry breaking to the maximal torus $U(1)^{n-1}$ with arbitrarily chosen holonomy eigenvalues $\mu_m$. 
In the situation of non-maximal breaking, 
some of the eigenvalues of the holonomy become equal, resulting in some monopoles acquiring zero mass. 
The form of the relative mass matrices defined as inverses suggests that dramatic things happen when one or more of the constituents acquire zero mass. However, as is clear from the explicit forms of $M, M_\mon$ in equations (\ref{eq:mmcal}, \ref{eq:relmmcal}, \ref{eq:mmmon}, \ref{eq:relmmmon}),
all limits can be taken smoothly. 
This assertion was explicitly checked for the trivial holonomy caloron, with all but one monopoles having zero mass. 
Therefore one can study most efficiently all symmetry breaking patterns, both for $k=1$ calorons and for monopoles of type $(1,1,\ldots1)$, just by inserting the proper values for $\mu_m$, rather than having to calculate the metric for each case separately. 
Consider, both for the caloron and for the $(1,1,\ldots,1)$ monopole, 
the situation of $N-1$ monopoles turning massless
\beq
\nu_K, \ldots, \nu_{K+N-2} = 0, \quad \mu_K = \ldots = \mu_{K+N-1}, \label{eq:massless}
\eeq
resulting in an enhanced residual symmetry to $SU(N) \times U(1)^{n-N}$.
The corresponding center of mass radii no longer appear in the expression for the action and energy densities \cite{MC}, as follows from
\beq
\prod_{m=1}^n A_m \rightarrow 
\left\{\prod_{m'=K+N-1}^n A_{m'}\right\}
\left(\bea{cc} r_{K-1} &R_c \\0& r_{K+N-1} \eea\right) 
\left(\bea{cc} c_{K-1}& s_{K-1}\\s_{K-1}&c_{K-1}\eea \right)
\frac{1}{r_{K-1}}
\left\{\prod_{m=1}^{K-2} A_m \right\}
\eeq
Here
\beq
R_c = | \vec\rho _K| + \ldots + | \vec\rho _{K+N-1}| = \pi \tr_2 \sum_{m = K}^{K+N-1} \zeta^\dagger_{(m)} \zeta^{\vphantom \dagger}_{(m)}
\label{eq:cloud}
\eeq
denotes what is known in the monopole literature as the "non-abelian cloud" parameter \cite{LeeWeiYi2}. 
It is seen from the right hand side of \refeq{cloud} that it is $SU(N)$ invariant. 
From the ADHM-Nahm construction (\ref{eq:adhmzm},\ref{eq:adhmveld}), this $SU(N)$ symmetry is seen to leave the holonomy invariant. 
It will descend to the quotient in the \hkqc\ of the metric, and therefore, the metric will be $SU(N)$ invariant as well, much like in the case of the trivial holonomy caloron. 
As the explicit form of the metric can readily be found by inserting \refeq{massless} in the mass matrices (\ref{eq:mmmon}, \ref{eq:mmcal}), it will not be given here. 
The $SU(N)$ transformations mixes the positions of the massless monopoles, which therefore do not exist as individual particles. 
A way of seeing this physically is that the intrinsic length scales of the monopoles, proportional to their inverse masses, become infinitely large as their masses become small,
so that they overlap and lose their indentities. 
This appearance of massless particles and infinite length scales illustrates a very general feature of systems near a transition to a more symmetric phase. 

The fact that the $SU(n+1)$ $(1,1,\ldots,1)$ monopole and the $SU(n)$ $k=1$ caloron both consist out of $n$ constituent BPS monopoles in combination with the fact that the former can be obtained out of an $SU(n+1)$ caloron, suggests a great similarity between their metrics. 
We consider the relevant situation for quantum chromodynamics, the $SU(3)$ caloron. 
Removing one monopole to infinity gives the $SU(3)$ monopole of type (1,1). There remain two constituents, of masses proportional to $\nu_1, \nu_2$.
The relative metric of the $(1,1)$ monopole is Taub-NUT with positive mass parameter. 
\beq
\label{eq:tn}
g_{TN} = U ( \vec\rho ) d \vec\rho ^2 + U ( \vec\rho )^{-1} (\frac{d \psi}{4 \pi} + \frac{ \vec \w(\vec\rho)}{4\pi} \cdot d \vec\rho)^2, \quad U(\vec\rho) = \frac{\nu_1 \nu_2}{\nu_1 + \nu_2} + \frac {Q}{4 \pi |\vec\rho|},
\eeq
$\vec\rho$ denoting the separation of the constituents, $Q = 1$. 
The relative metric for the $SU(2)$ caloron is also a Taub-NUT \cite{PLB} \cite{NPB}. (The metric obtained there checks with \refeq{tn} apart from the normalisation $4\pi^2$, as $\pi\rho^2, \Upsilon$ in \cite{NPB} corresponds to $|\vec r|, \upsilon$ in \refeq{tn}). 
However, the interaction strength, depending on the distance between the monopoles, for the caloron is $Q=2$, twice that of the $SU(3)$ monopole. 
Both solitons can be considered as built out of two interacting constituent BPS monopoles, and have a four dimensional relative moduli space. 
Each matching point in the Nahm construction gives rise to an interaction between monopoles of distinct type, 
this is to be expected. 
The $SU(3)$ $(1,1)$ monopole has one matching point, at $z = \mu_2$ whereas the $SU(2)$ caloron has one additional at $ z = \mu_1 + 1$ to close the circle, in the situation of two constituents equal to the other. 
In \cite{LeeLu} this was attributed to the fact that the constituent monopoles in the $SU(3)$ (1,1) case are charged with respect to different $U(1)$, whereas for the caloron, they are oppositely charged with respect to the {\em same} $U(1)$, generated by $\vec \omega\cdot \vec\tau$. 

In conclusion, we have presented results for the metric on moduli spaces in a 
unified description that incorporates instantons, calorons and monopoles.

\section{Acknowledgements}
Pierre van Baal is gratefully acknowledged for discussions and critically reading earlier versions of the manuscript.
Conversations and correspondence with Conor Houghton have been very stimulating. This work was financially supported by a grant from the FOM/SWON 
Association for Mathematical Physics. 

\section*{Appendix}
{\large\bf The $(1,1,\ldots, 1)$ monopole}\\
The Nahm construction of the $(1,1,\ldots,1)$ monopole is similar to that of the $k=1$ $SU(n)$ caloron. The main difference is that the circle is replaced by the interval $[\mu_1, \mu_n]$. For the $\oo$ monopole, the singularities reside at $z = \mu_2, \ldots, \mu_{n-1}$ \cite{NahAll, HurMur, WeiYi}. Like for the caloron we introduce $\Delta^\dagger=(\lambda^\dagger(z), \frac{1}{2 \pi i} \hat D^\dagger_x(z))$,
\beq
\lambda(z) = \sum_{m=2}^{n-1} \delta (z - \mu_m) \zeta_m, \quad
\hat D_x(z)=\sigma_\mu\hat D_x^\mu(z)=\ddz+\hat A(z)-2\pi ix,
\eeq
where $\hat A(z)$ is now defined on $[\mu_1, \mu_{n}]$. The Nahm construction is performed in terms of the normalised zero modes $v(x)$ of $\Delta(x)$ 
\beqa
&&\quad v(x) = \left( \bea{c} s_x \\ \hat \psi_x(z) \eea \right), \quad \frac{1}{2 \pi i} \hat D^\dagger_x(z) \hat \psi^m_x(z) + \sum_{m'=2}^{n-1} \delta(z - \mu_{m'}) \zeta^\dagger_{m'} s^m_{x m'} =0,\\
&&\quad v^\dagger(x) v(x) = s^\dagger_x s_x +\int_{\mu_1}^{\mu_n} dz \hat \psi^\dagger_x(z) \hat\psi_x(z) = 1_n, \label{eq:minp}
\eeqa
where $\hat \psi_x(z) = (\psi_x^1(z), \ldots, \psi_x^n(z))$ contains the $n$ two-spinors defined on the interval $[\mu_1, \mu_n]$, 
and $s \in \complex^{n-1,n}$. (The equation for $\hat\psi^m_x(z)$ is readily seen to have $n$ solutions for fixed $s_x$ \cite{HurMur}).
Though the monopole is a static solution, it is preferable to have $x_0$ included as a dummy variable, the $x_0$ dependence trivially being implemented by $v(x) = e^{2 \pi i x_0 z} v(\vec x)$, so as to write concisely
\beq
A_\mu(\vec x) = (\Phi(\vec x), \vec A(\vec x)) = v^\dagger(x) \partial_\mu v(x)
\label{eq:monahm}
\eeq
with the inner product defined as in \refeq{minp}. Performing all monopole calculations in terms of $\Delta(x)$ and $v$, the caloron formalism can be copied. 
In particular, it follows that for \refeq{monahm} to be \sd, $\Delta^\dagger(x) \Delta(x)$ should commute with the quaternions. This is equivalent to the monopole Nahm equation 
\beq
\ddz \hat A_j(z) = 2 \pi i \sum_{m=2}^{n-1} \delta(z - \mu_m) \rho^j_m.
\eeq
Its solution $\hat A_j(z)$ can be written in terms of $n-1$ position vectors $\vec y_m$, $\vec\rho_m = \vec y_m - \vec y_{m-1}$, comprised in $\vec y_\mon = ( \vec y_1, \ldots, \vec y_n)^t$,
\beq
\vec\rho_\mon= S^t_\mon \vec y_\mon \label{eq:rnahmmon}
\eeq
implying
\beq
\hat A_j = 2 \pi i \sum_{m=1}^{n-1} \chi_{[\mu_m, \mu_{m+1}]}(z) \vec y_m.
\eeq
Like for the caloron, there is a gauge action on the Nahm data
\beq
\hat A(z) \rightarrow \hat A(z) +i \ddz h(z), \quad\zeta_m \rightarrow \zeta_m e^{ih(\mu_m)}, \quad m = 2, \ldots, n-1 \label{eq:monga}
\eeq
with gauge group $\hat \cG_\mon = \{ g(z) | g: z \rightarrow e^{- i h(z)} \in U(1), g(\mu_1) = g(\mu_n) = 1\}$. The condition at the endpoints is required for $i d/dz$ to be hermitean on the space of gauge functions. 
Hence, for the monopole $\hat \cG_\mon = \{ g(z) | g: z \rightarrow e^{- i h(z)} \in U(1), g(\mu_1) = g(\mu_n) = 1\}$. The $\cG_\mon$ action can be used to set $\hat A_0(z)$ constant, and to undo the $U(1)$ phase ambiguities in relating $\zeta_m$ to $ \vec\rho _m, m = 2, \ldots, n-1$, hence $\zeta_m$ can be considered to have fixed phase. The monopole Nahm data can then be expressed in terms of $n-1$ quaternions 
\beq
\hat A_\mon = ( \hat A_1, \ldots, \hat A_{n-1})^t = 2 \pi (N_\mon^{-1} \frac{\tau_\mon}{4 \pi} + \vec y_\mon \cdot \vec\sigma), 
\eeq
$i\hat A_{\mon, m}$ denoting the value $\hat A(z)$ takes on $(\mu_m, \mu_{m+1})$.

In the gauge with constant $\hat A_0(z)$, the Green's function $f_x$ in the monopole Nahm construction is the solution to the differential equation 
\beq
\left\{\!\!\left(\!\frac{1}{2\pi i}\ddz \!-\! x_0\right)^{\!\!2} + \sum_{m=2}^{n-1} \chi_{[\mu_m, \mu_{m+1}]}(z) \, r_m^2 + \frac{1}{2 \pi}
 \sum_{m=2}^{n-1} \delta(z \!-\! \mu_m) |\vec y_m \!-\! \vec y_{m-1}| \!\right\}\!\!\hat f_x(z,z') = \delta(z-z').\label{eq:mondef}
\eeq
whereas transformations to other gauges are realised by \beq
\hat f_x(z,z') \rightarrow g(z) \hat f_x(z,z') g(z')^*, \quad g(z) \in \hat \cG_\mon . \label{eq:mongaf}
\eeq
The boundary condition for the monopole Green's function is determined by the requirement that $i \ddz$ be a hermitean operator, therefore the eigenfunctions of the left hand side of \refeq{mondef} vanish in the endpoints. This imposes by standard Sturm-Liouville theory
\beq
\hat f(\mu_1, z') = \hat f(\mu_n, z') = 0
\eeq
for the Green's function. 
This boundary condition is automatically satisfied when obtaining the monopole Green's function from the caloron Green's function, taking the limit $|\vec y_n| \rightarrow \infty$. The $x_0$ dependence of the monopole Green's function is trivial \beq
\hat f_x(z,z') = e^{2 \pi i x_0 ( z - z')} \hat f_{\vec x}(z,z').
\eeq
The metric on the monopole moduli space is determined in terms of the $L_2$ norm of gauge orthogonal solutions $Z_\mon$ to the linearised Bogomol'nyi equations. 
With $A_0$ identified as the Higgs field, and assuming all fields and zero modes being static, the conditions for a tangent vector to the monopole moduli space are identical to those for on the tangent vector to an instanton moduli space,
hence $Z_\mon$ satisfies
\beq
D^{{\rm ad}^\dagger}(A) Z_\mon = 0,
\eeq
where $\partial_0$ acts trivially, but is kept to make later derivations more transparant. Metric and K\"ahler forms read
\beq
(g, \hkom)(Z_\mon,Z'_\mon) = \frac{1}{4 \pi^2}\int_{\re^3} d^3x \Tr Z^\dagger_\mon(\vec x) Z'_\mon(\vec x) \label{eq:monmet}.
\eeq
The formalism to compute the metric is copied from the caloron case. A tangent vector to the monopole moduli space is given by
\beq
Z_{\mon \mu}(\vec x) =\int_{[\mu_1,\mu_n]^2} dz dz' \left( \sum_{m'=2}^{n-1} s^\dagger_x \hat c_{m'} \delta(z - \mu_{m'}) + \hat \psi_x(z)\right) \hat f_x(z,z') \sigma^\dagger_\mu \hat \psi_x(z') - h.c.
\eeq
in terms of a tangent vector to the moduli space of monopole Nahm data
\beq
C = \left( \bea{c} \hat c(z)\\ \hat Y(z) \eea \right),\quad \hat c(z) = \sum_{m=2}^{n-1} \hat c_m \delta(z - \mu_m), \quad 
\eeq
satisfying the deformation and gauge orthogonality equations
\beqa
\ddz \hat Y_i(z) &=& - i\pi \tr_2 \sum_{m=2}^{n-1} \bar \sigma^i (\zeta^\dagger_m \hat c^{\vphantom \dagger}_m + \hat c^\dagger_m \zeta^{\vphantom \dagger}_m) \delta (z - \mu_m),\nonumber \\
\ddz \hat Y_0(z) &=& -i \pi \sum_{m=2}^{n-1} \tr_2 ( \zeta_m^\dagger \hat c_m - {\hat c}^\dagger_m \zeta_m)\delta (z - \mu_m). \label{eq:mongo}
\eeqa
To derive the analogue for monopoles of Corrigan's formula we trade each matrix multiplication in \refeq{cor} for an integration over $[\mu_1, \mu_2]$ or an inner product of type (\ref{eq:minp}) and use the trivial $x_0$ dependence of $v(x)$ and $f_x(z,z')$ for the monopole to obtain
\beqa
\Tr Z^\dagger_\mon(x) Z'_\mon(x)\!\!\!& =&\!\!\! - \nabla^2 \int_{[\mu_1, \mu_2]} dz \left( [ \hat Y^\dagger(z) \hat Y'(z) + {\hat c}^\dagger (z) < \hat c'> ] \hat f_x(z,z) \right) \label{eq:moncor}\\
\!\!\!&+&\!\!\! \half \nabla^2 \int_{[\mu_1, \mu_2]^2} dz dz' \left([\cC(z) + \cY(z)] \hat f_x(z,z') [\cY_x'^{\dagger}(z') + \cC'^{\dagger}(z')] \hat f_x(z',z)\right)\nonumber,
\eeqa 
with $\cC(z) = \sum_{m=2}^{n-1} \hat c^\dagger_m \zeta_m \delta(z - \mu_m)$, $\cY_x(z) = ( 2 \pi i)^{-1} \hat Y^\dagger(z) \hat D_x(z)$. 
The monopole metric is evaluated from eqs. (\ref{eq:monmet}, \ref{eq:moncor}) by partial integration, along the lines of the derivation in section \ref{sec:isomadhmn}. 
The monopole Green's function $f_x(z,z')$ behaves as in \refeq{fas}. 
Thus we arrive at the isometric property of the Nahm construction for $\oo$ monopoles,
\beq
(g,\hkom)_{\cal M}(Z_\mon,Z'_\mon) = \Tr 
\left(<\hat Y^\dagger Y' > + 2< \hat c^\dagger> < \hat c'> \right), 
\quad
 <\!H\!> \equiv \int_{[\mu_1, \mu_n]} H(z) dz.
\label{eq:monmetev}
\eeq
An infinitesimal gauge transformation $\delta \hat X(z)$ is applied to obtain gauge orthogonality of the tangent vector $C$
\beqa 
&&\label{eq:montv}
\hat c(z)= \sum_{m=2}^{n-1} \delta(z - \mu_m) \hat c=\sum_{m=2}^{n-1} \delta(z - \mu_m) \left(\delta \zeta_m +i \zeta_m \delta \hat X(\mu_m)\right), \\
&&\hat Y(z)= i\sum_{m=1}^{n-1}\chi_{[\mu_m, \mu_{m+1}]} \hat Y_m = \frac{1}{2 \pi i} \left(\delta \hat A(z) + i\ddz \delta \hat X(z) \right). \nonumber 
\eeqa
It vanishes in the endpoints $z = \mu_1, z = \mu_n$ and satisfies
\beq
\!-\! \frac{1}{2 \pi} \frac{d^2 \delta \hat X(z)}{dz^2} + 2 \delta \hat X(z) \!\sum_{m=2}^{n-1} \delta(z \!-\! \mu_m) |\vec\rho_m|\! =\! \!\sum_{m=2}^{n-1} \delta( z\! -\! \mu_m) \!
\left[ \frac{d \ttau_m}{4 \pi \nu_m} \!-\! \frac{d \ttau_{m-1}}{4 \pi \nu_{m-1}} - |\vec\rho_m|\vec\omw_m(\vec\rho_m) \!\cdot \! d \vec\rho_m \!\right]\!.
\eeq
Therefore, it is piecewise linear and fixed by $\delta \hat X = (\delta \hat X_2, \ldots, \delta \hat X_{n-1})^t$, $\delta \hat X_m = \delta \hat X(\mu_m)$, $m = 2, \ldots, n-1$ where
\beq
\frac{1}{2\pi}(S^t_\mon N^{-1}_\mon S_\mon + V^{-1}_\mon) \delta \hat X = 
(S^t_\mon N^{-1}_\mon \frac{d \ttau_\mon}{4 \pi} - V^{-1}_\mon \vec W_\mon S^t_\mon \cdot d \vec y_\mon),
\eeq
(see eqs. (\ref{eq:mmmon}, \ref{eq:sm}) for definitions).
With the compensating gauge function found, the remaining manipulations to retrieve the uncentered monopole metric in \refeq{lwyu} from eqs. (\ref{eq:monmetev}, \ref{eq:montv}) differ only in the $\mon$ label and the dimensions of the matrices from those in section \ref{sec:caldir} and are therefore not repeated here. 

To compute the metric using the \hkqc\ we follow and summarise the reasoning in \cite{MurNot, GibRycGot} and section \ref{sec:hkqccm}. 
We have to find the metric on $\cN_\mon/{\hat \cG}_\mon$, where $\cN_\mon$ is the 
subset of the space $\hat {\cal A}_\mon$ of monopole Nahm data containing the solutions to the Nahm equations. 
Making use of the $U(1)$ gauge symmetry for monopole in \refeq{monga}, we can restrict ourselves to piecewise constant $\hat A(z)$, characterised by $n-1$ quaternions corresponding to its values on the subintervals. Together with the $n-2$ complex two vectors giving the matching data, form the space ${\hat{\cal A}}_{0 \mon} = \qu^{n-1} \times \complex^{n-2, 2} \ni (y_\mon, \zeta_\mon)$. This space has natural metric and \K\ forms
\beq
(g, \hkom) = d y^\dagger_\mon N_\mon \otimes d y_\mon + 2 d \zeta^\dagger_\mon \otimes d\zeta_\mon \label{eq:monst}
\eeq
The set of piecewise constant solutions to the Nahm equations form $\cN_{0, \mon}$, which is a subset of ${\hat{\cal A}}_{0 \mon}$. The vector part of a piecewise constant solution to the monopole Nahm equation (\ie $\cN_{\mon, 0}$) is fixed by \refeq{rnahmmon}.
We introduce the phases of $\zeta_\mon$ as $\psi_\mon = (\psi_2, \ldots, \psi_{n-1})^t$. Having gauge fixed to constant $\hat A(z)$, 
the residual $U(1)$ gauge symmetry consists of gauge functions having piecewise linear and continuous logarithms, which vanish in the endpoints $z = \mu_1$ and $z = \mu_m$. 
This results in an $\re^{n-2}$ action on ${\hat{\cal A}}_{0 \mon}$, characterised by
\beq
y_\mon \rightarrow y_\mon - \frac{1}{2\pi} N_\mon^{-1} S_\mon t_\mon, \quad \psi_\mon \rightarrow \psi_\mon + 2 t_\mon, \quad t_\mon\in\re^{n-2}, \label{eq:mac}
\eeq
with moment map, zero set and invariants given by
\beq
\mmu_\mon = - \frac{1}{2 \pi} S^t_\mon \vec y_\mon + \frac{ \vec\rho _\mon}{2\pi},\quad 
\vec\rho_\mon = S^t_\mon \vec y_\mon
, \quad
\tau_\mon = 4 \pi N_\mon y_{0 \mon} + S_\mon \psi_\mon.
\eeq
Having established a suitable notation, the algebra to obtain the metric and \K\ forms for the uncentered monopole in \refeq{lwyu} is now nearly identical to the \hkqc\ of the uncentered caloron metric, and one readily retrieves \refeq{lwyu}. Actually, one only has to insert the $\mon$ labels at appropriate places, 
just realising that the dimensionalities of the objects are slightly different. 
\end{document}